\documentclass{aa}  
\usepackage{amsmath,amssymb}
\usepackage{graphicx}
\usepackage{txfonts}

\begin{document}
\title{The Spatial Range of Conformity} 
\author{Martin Kerscher} 
\institute{Ludwig--Maximilians Universtät München, 
  Fakultät für Physik, Schellingstr. 4, D-80799 München\\
  \email{martin.kerscher@lmu.de}} 

\date{March 5, 2018} 

\abstract
{Properties of galaxies like their absolute magnitude and their
  stellar mass content are correlated.  These correlations are tighter
  for close pairs of galaxies, which is called galactic conformity.
  In hierarchical structure formation scenarios, galaxies form within
  dark matter halos.  To explain the amplitude and the spatial range
  of galactic conformity two--halo terms or assembly bias become
  important.  }
{With the scale dependent correlation coefficients the amplitude and
  the spatial range of conformity are determined from galaxy and halo
  samples. }
{The scale dependent correlation coefficients are introduced as a new
  descriptive statistic to quantify the correlations between
  properties of galaxies or halos, depending on the distances to other
  galaxies or halos.  These scale dependent correlation coefficients
  can be applied to the galaxy distribution directly. Neither a
  splitting of the sample into subsamples, nor an a priori clustering
  is needed. }
{This new descriptive statistic is applied to galaxy catalogues
  derived from the Sloan Digital Sky Survey~III and to halo
  catalogues from the MultiDark simulations.  In the galaxy sample the
  correlations between absolute Magnitude, velocity dispersion,
  ellipticity, and stellar mass content are investigated.  The
  correlations of mass, spin, and ellipticity are explored in the halo
  samples.
  Both for galaxies and halos a scale dependent conformity is
  confirmed.  Moreover the scale dependent correlation coefficients
  reveal a signal of conformity out to 40\,Mpc and beyond.  The halo
  and galaxy samples show a differing amplitude and range of
  conformity.}
{}


\maketitle

\section{Introduction}

The clustering of galaxies in space is an important observational
constraint for models of structure formation in the Universe.  Often
galaxies are treated as points in space and one compares the
clustering properties of this point distribution to models of
structure formation.  However galaxies are extended objects and come
in different flavours.  Their properties are categorised and
quantified. One considers the luminosity, the shape, the substructure
or spectroscopic features of a galaxy, to name only a few.
As an extension, galaxies are still treated as points, but the
properties of the galaxies are assigned to the points as marks.
This establishes at each position of a galaxy a multidimensional
space. Depending on the physical problem, different methods for the
analysis of such a marked point set have been devised:

The concept of bias was developed to account for the stronger
clustering of galaxy--clusters compared to the clustering of galaxies
themselves \citep{kaiser:onspatial,bardeen:gauss}. Currently bias is
often used to describe the differences between the clustering of
luminous and dark matter (see \citealt{desjacques:largescale} for a
recent review)

With luminosity-- and morphology--segregation one describes the
differences in the spatial clustering of dim versus luminous galaxies,
of early--type (e.g.\ ellipticals) versus late--type galaxies (e.g.\
spirals), or of red versus blue galaxies, etc.\
\citep{ostriker:luminosity, hamilton:evidence, willmer:southern}.  In
most cases the ratios of the two--point correlation functions,
determined from sub--samples of the galaxy distribution are used to
quantify these segregation effect (see e.g.\ \citealt{zehavi:sdssfinal}).

The morphology density relation indicates that early type galaxies
tend to reside in more dense environments compared to late type
galaxies.  There are numerous observations confirming this
\citep{dressler:galaxy, postman:morphology-density,
  andreon:morphologyIII, vanderwell:physical}. Effects of the
morphology density relation are typically confined to groups and
clusters of galaxies (see however \citealt{binggeli:abundance}).

Conformity is an expression from sociology, it is the act of matching
attitudes and behaviours to group norms.  With galactic conformity one
is investigating how strongly the properties of galaxies conform with
each other, if they are located in a group around a bright dominating
galaxy or in a dark matter halo \citep{weinmann:properties}.
Galactic conformity is typically quantified by first determining the
central galaxy within a group of galaxies. Then e.g.\ the fraction of
late type galaxies in the cluster is plotted against the mass of the
group depending on the type of the central galaxy.  Hence, galactic
conformity is an extension of the morphology density relation, with
the focus on the bright central galaxy as the determinant for the
galactic properties. Not only the types but also the
colours, the star formation rates, or other properties of the galaxies
are being used.
\cite{kauffmann:reexamination} plotted the fraction of star forming
galaxies against the (projected) distance from the central galaxy,
showing that conformity is scale dependent, at least on small scales.
In hierarchical structure formation scenarios, galaxies form within
dark matter halos.  To explain the amplitude and the spatial range of
galactic conformity two--halo terms or assembly bias becomes
important.  Using the halo model \cite{hearin:beyond} were able to
model such a scale dependence using 2--halo conformity from assembly
bias. A comparison of semi--analytic models reveals different patterns
in the scale dependence of halo conformity between the models (see the
discussion in \citealt{lacerna:galactic} and the references therein).
Quantitative scale dependent methods are needed to discriminate
these different approaches. This is especially important if one wishes to
quantify the influence of large--scale structures on the conformity.
Then one needs measures of conformity wich are also sensitive on large
scales.

As a new descriptive statistic based on mark correlation functions, the
scale dependent correlation coefficients are introduced to quantify
dependencies between properties of galaxies (or halos).
The scale dependent correlation coefficients measure the strength of
the correlations between the intrinsic properties of a galaxy and how
these correlations on one galaxy depend on the presence of another
galaxy at a distance of $r$ (similarly for halos).  Hence they allow a
scale dependent measurement of the \emph{conformity}.
To estimate the scale dependent correlation coefficients suitably
weighted pair counts of all the galaxies are used.  Conceptually this
is a major benefit, all pairs are counted. The galaxy sample is not
split into several parts, e.g.\ early type, late type, nor any
grouping into clusters is necessary. No new (nuisance) parameters are
introduced into the analysis.

In section~\ref{sec:method} the scale dependent correlation
coefficients are defined. They are used in section~\ref{sec:sdss} to
analyse galaxy samples from the Sloan Digital Sky Survey (SDSS), and
in section~\ref{sec:simulation} for halo samples from the MultiDark
dark matter simulations.  A summary and conclusion is given in
section~\ref{sec:summary}.  In Appendix~\ref{sec:samples} the
construction of the galaxy and halo samples is detailed, and a simple
toy model is presented in Appendix~\ref{sec:toy}.

\section{The method}
\label{sec:method}

The well known definitions of covariance and correlation coefficient
are reviewed in the next subsection.  This discussion serves as a
blue--print for the definition of the \emph{scale dependent
  correlation coefficient} in subsection~\ref{sec:corrscale}.
The definitions are given explicitly for galaxies with absolute
r--magnitude $M_r$ and ellipticity $e$.  For the SDSS galaxies and the
halo samples from the MultiDark simulations, also other properties are
used as marks in the analysis below (see Appendix~\ref{sec:samples}
for details).
In the following the positions of the galaxies together with their
properties are interpreted as a realisation of a marked point process
\citep{beisbart:wuppertal}. The two-point theory of marked point
processes was developed by \citet{stoyan:oncorrelations} and is nicely
reviewed in \citet{stoyan:fractals}.
First applications of mark correlation function  to
galaxy samples are discussed in
\cite{beisbart:luminosity,szapudi:correlationspscz,beisbart:wuppertal}
and to halo simulations in
\cite{gottloeber:merger,faltenbacher:halos,sheth:environmental}.

\subsection{Correlations between properties of galaxies or halos}

In this subsection only the intrinsic properties of galaxies or halos
will be of interest, irrespective of their position in space.  The
joint probability densities provides a suitable tool to describe the
statistics of the galaxy (or halo) properties.
${\cal M}(M_r,e)$ is the probability density of finding a galaxy with
absolute r-magnitude $M_r$ {\em and} with ellipticity $e$ in our
sample.  Marginalising ${\cal M}(M_r,e)$, one obtains the probability
density of the ellipticity
${\cal M}(e)=\int{\rm d} M_r {\cal M}(M_r,e)$ and similarly the
probability density of the absolute r--magnitude ${\cal M}(M_r)$.
The moments are defined in the usual way. E.g.\ the $k$th--moment of
the ellipticity--distribution is
$\overline{e^k} = \int{\rm d} e\ {\cal M}(e)\ e^k;$ with the mean
ellipticity $\overline{e}=\overline{e^1}$ and the variance
$\sigma_e^2=\overline{e^2}-\overline{e}^2$.
If $M_r$ and $e$ are independent
${\cal M}(M_r,e)={\cal M}(M_r){\cal M}(e)$, however in general this is
not the case. To quantify the dependency the covariance and
correlation coefficient of $M_r$ and $e$ are used. The covariance is
defined as
\begin{equation}
\label{eq:cov-onepoint}
  \text{cov}(M_r,e)
  = \int{\rm d} e \int{\rm d} M_r\ {\cal M}(M_r,e)\ 
  \left(M_r-\overline{M_r}\right) \left(e-\overline{e}\right) .
\end{equation}
Suitably normalised one obtains the well known correlation coefficient
\begin{equation}
\label{eq:cor-onepoint}
\text{cor}(M_r,e)=\frac{\text{cov}(M_r,e)}{\sigma_{M_r}\ \sigma_e}.
\end{equation}
By definition $-1\le \text{cor}(M_r,e) \le 1$.  The larger the modulus
of $\text{cor}(M_r,e)$, the stronger the (anti-)\,correlation between
$M_r$ and $e$.

\subsection{The scale dependent correlation coefficient}
\label{sec:corrscale}

Calculating the above defined correlation coefficients under the
condition that another galaxy is at a distance of $r$ one arrives at
the desired statistic describing scale dependent correlations.  To
define these scale dependent correlation coefficients the flexible
framework of mark correlation functions is used
\citep{stoyan:oncorrelations,beisbart:luminosity}.

$\varrho_1(\mathbf{x},M_r,e)$ is the probability density of finding a
galaxy at $\mathbf{x}$ with an absolute magnitude $M_r$ and an
ellipticity $e$.  For a homogeneous point distribution this splits
into $\varrho_1(\mathbf{x},M_r,e)=\varrho\,{\cal M}(M_r,e)$ where
$\varrho$ denotes the mean number density of galaxies in space and
${\cal M}(M_r,e)$, the already defined probability density of finding a
galaxy with absolute r--magnitude $M_r$ and ellipticity $e$.
Slightly extending the notation from above, $M_{r,i}$ and  $e_i$ are 
the absolute r--magnitude and ellipticity of the galaxy at the
position $\mathbf{x}_i$.
Accordingly,
$\varrho_2\left((\mathbf{x}_1,M_{r,1},e_1),(\mathbf{x}_2,M_{r,2},e_2)\right)$
quantifies the probability density of finding two galaxies at
$\mathbf{x}_1$ and $\mathbf{x}_2$ with the absolute magnitudes
$M_{r,1}$, $M_{r,2}$ and the ellipticities $e_1$, $e_2$, respectively.
For an isotropic and homogeneous point set $\varrho_2(\cdots)$ only
depends on the separation $r=|\mathbf{x}_2-\mathbf{x}_1|$ and the
spatial product density is then given by $\varrho^2\,(1+\xi_2(r))$,
with the well known two--point correlation function $\xi_2(r)$.

It is useful to consider the conditional mark probability density
defined as
\begin{gather}  
\label{eq:cond-mark-density}
{\cal M}_2\left(M_{r,1},e_1,M_{r,2},e_2\,|\,r\right)=
\frac{\varrho_2\left((\mathbf{x}_1,M_{r,1},e_1),(\mathbf{x}_2,M_{r,2},e_2)\right)}
{\varrho^2(1+\xi_2(r))}.
\end{gather}
${\cal M}_2\left(M_{r,1},e_1,M_{r,2},e_2\,|\,r\right)$ is the
probability density of the absolute magnitudes $M_{r,1}$, $M_{r,2}$,
and ellipticities $e_1$, $e_2$ under the condition that this pair of
galaxies is separated by $r=|\mathbf{x}_1-\mathbf{x}_2|$.
We speak of \emph{mark--independent} clustering, if
${\cal M}_2\left(M_{r,1},e_1,M_{r,2},e_2\,|\,r\right)={\cal
  M}(M_{r,1},e_1){\cal M}(M_{r,2},e_2)$ factorises and does not depend
on the pair separation $r$. In such a case the absolute magnitudes and
the ellipticities of galaxy pairs with a separation $r$ are not
different from any other pair of galaxies. On the contrary,
mark--dependent clustering or \emph{mark segregation} implies that the
marks on certain galaxy pairs show deviations from the global mark
distribution.

The conditional probability density
${\cal M}_2\left(M_{r,1},e_1,M_{r,2},e_2\,|\,r\right)$ is used to calculate
the scale--dependent correlation coefficient:
\begin{multline}  
\label{eq:cor}
\textbf{cor}\left(M_r,e\,|\,r\right) = \\
 \iiiint\limits_{M_{r,1},e_1,M_{r,2},e_2}\ 
{\cal M}_2\left(M_{r,1},e_1,M_{r,2},e_2\,|\,r\right)
\frac{ \left(M_{r,1}-\overline{M_r}\right)(e_1-\overline{e}) }
     { \sigma_{M_{r}}\, \sigma_e },
\end{multline}
with the abbreviation 
\[
\iiiint\limits_{M_{r,1},e_1,M_{r,2},e_2} 
:= \int{\rm d} M_{r,1}\int{\rm d} e_1 \int{\rm d} M_{r,2}\int{\rm d} e_2 .
\]
Only the correlation coefficient between $M_r$ and $e$ on galaxy~1 is
calculated, the marks on galaxy~2 are integrated out.  One should
compare this definition with Eqs.~(\ref{eq:cov-onepoint})
and~(\ref{eq:cor-onepoint}) to see the close analogy.
$\textbf{cor}\left(M_r,e\,|\,r\right)$ quantifies the correlation
between the absolute magnitude $M_r$ and the ellipticity $e$ on
\emph{one} galaxy under the condition that another galaxy is at a
distance of $r$. If there is an environmental dependency one expects
$\textbf{cor}\left(M_r,e\,|\,r\right)\ne \text{cor}(M_r,e)$. For large
separations the environmental dependency has to vanish and one gets
$\textbf{cor}(M_r,e\,|\,r\rightarrow\infty)= \text{cor}(M_r,e)$.


Similar to Eq.\,\ref{eq:cor} one can define the scale dependent mean 
\begin{equation}  
\label{eq:meanr}
\overline{e}(r) = \iiiint\limits_{M_{r,1},e_1,M_{r,2},e_2}
{\cal M}_2\left(M_{r,1},e_1,M_{r,2},e_2\,|\,r\right) \, e_1
\end{equation}
and with $\overline{e}(r)$ the scale dependent variance
\begin{equation}  
\label{eq:varr}
\sigma_e^2(r) = \iiiint\limits_{M_{r,1},e_1,M_{r,2},e_2}
{\cal M}_2\left(M_{r,1},e_1,M_{r,2},e_2\,|\,r\right) \  (e_1 -\overline{e}(r))^2 .
\end{equation}
The scale dependent mean $\overline{e}(r)$ and and the scale dependent
variance $\sigma_e^2(r)$ are the mark correlation functions $k_m()$
and var() as defined in \citet{beisbart:luminosity}.
The scale dependent mean and variance allow the definition of
an alternative scale dependent correlation coefficient\footnote{This
  alternative definition of
  $\widetilde{\textbf{cor}}\left(\cdot,\cdot|\,r\right)$ was suggested
  to me by Simon White after the first submission of this article.}
\begin{multline}  
\label{eq:corrtilde}
\widetilde{\textbf{cor}}\left(M_r,e\,|\,r\right) = 
\iiiint\limits_{M_{r,1},e_1,M_{r,2},e_2} 
{\cal M}_2\left(M_{r,1},e_1,M_{r,2},e_2\,|\,r\right) \times \\
\times \frac{ \left(M_{r,1}-\overline{M_r}(r)\right)(e_1-\overline{e}(r)) }
     { \sigma_{M_{r}}(r)\, \sigma_e(r) }.
\end{multline}
This defines the correlation coefficient relative to the mean and
variance of galaxies with another galaxy at a distance of $r$ (cf.\
equation~(\ref{eq:cor})).
In Appendix~\ref{sec:toy} both $\textbf{cor}()$ and
$\widetilde{\textbf{cor}}()$ are calculated for a simple toy model 
with a built in scale. With $\textbf{cor}()$ the scale can be detected 
easily from the samples, whereas $\widetilde{\textbf{cor}}()$ is not 
depending on the built in scale.
As another example consider
$\widetilde{\textbf{cor}}\left(M_r, m_\text{st}|\,r\right)$ with
$r\in[1,3]$\,Mpc, the correlation coefficient between $M_r$ and
$m_\text{st}$ of all the galaxies with another galaxy at a distance
$r\in[1,3]$\,Mpc.  Then
$\widetilde{\textbf{cor}}\left(M_r,m_\text{st}|\,r\right)/
\text{cor}\left(M_r,m_\text{st}\right)$
quantifies the deviation from the corresponding correlation
coefficient of all galaxies as visible in Fig.\,\ref{fig:sdsshists}
below.

It is straightforward to estimate mark correlation functions like
$\textbf{cor}\left(M_r,e\,|\,r\right)$ from a galaxy catalogue. The
basic idea derives from eqs.(\ref{eq:cond-mark-density},
\ref{eq:cor}): one adds up every pair $(i,j)$ of galaxies separated by
$r$ weighted by
$\left(M_{r,i}-\overline{M_r}\right)(e_j-\overline{e})$. Then one
divides by the number of pairs with separation $r$. Suitably
normalised one obtains an estimate of
$\textbf{cor}\left(M_r,e\,|\,r\right)$.  Analog ideas apply for the
estimation of $\widetilde{\textbf{cor}}\left(M_r,e\,|\,r\right)$.  A
more detailed discussion and a comparison of several estimators for
mark correlation functions is given in the Appendix of
\citet{beisbart:luminosity}.

The procedure offers a built-in significance test
\citep{beisbart:luminosity, grabarnik:correct}. One can redistribute
the galaxy properties within the sample randomly, holding the galaxy
positions fixed.  In that way one mimics a galaxy distribution with
the same spatial clustering and the same one-point correlations
$\text{cor}(M_r,e)$, but without any environmental dependency of these
correlations. Given the original data set, such samples with
mark--independent clustering can be simulated easily and the
fluctuations around
$\textbf{cor}\left(M_r,e\,|\,r\right)/\text{cor}(M_r,e)=1$ can be
quantified.

\section{Scale dependent correlation coefficients of 
galaxies from the SDSS DR12}
\label{sec:sdss}

The Sloan Digital Sky Survey (SDSS), data release 12 (DR12) includes a
magnitude limited sample of galaxies, the main galaxy catalogue
\citep{alam:dr12,eisenstein:sdssIII}. For these galaxies photometric
and spectroscopic, as well as derived properties are available from
the SDSS database.
The scale dependent correlation coefficients are estimated from volume
limited samples constructed from the main galaxy catalogue.  The
extinction and K--corrected absolute magnitude $M_r$, the two
dimensional ellipticity $e$ on the sky, the spectrally determined
velocity dispersion $\sigma_v$, and the logarithmic stellar mass
$m_\text{st}$ are assigned to each of the galaxies as marks.
The construction of the volume limited samples and details on the
estimation and normalisation of the marks $M_r$, $e$, $\sigma_v$, and
$m_\text{st}$ are given in Appendix~\ref{sec:samples-sdss}.

Besides introducing the scale dependent correlation coefficients as a
descriptive statistic for measuring conformity, the focus in this
article is on the spatial range of conformity, i.e.\ from how far out the
correlations between properties on one galaxy are influenced.
The absolute magnitude, the stellar mass content, the velocity
dispersion and ellipticity have been chosen as marks, because they show
appreciable correlations already for the whole sample (see
Table\,\ref{tab:sdsscorrmatv600}). The legitimate expectation is that a
scale dependence of conformity can be resolved easily for these marks.
With the absolute magnitude, the velocity dispersion and the stellar
mass content different aspects of the unobservable overall mass of the
galaxy are investigated. The ellipticity is used as a tracer of the
shape of the galaxy.
In the halo samples below analog parameters were chosen as marks.
%

\begin{figure*}
  \centering
  \includegraphics[width=5.5cm]{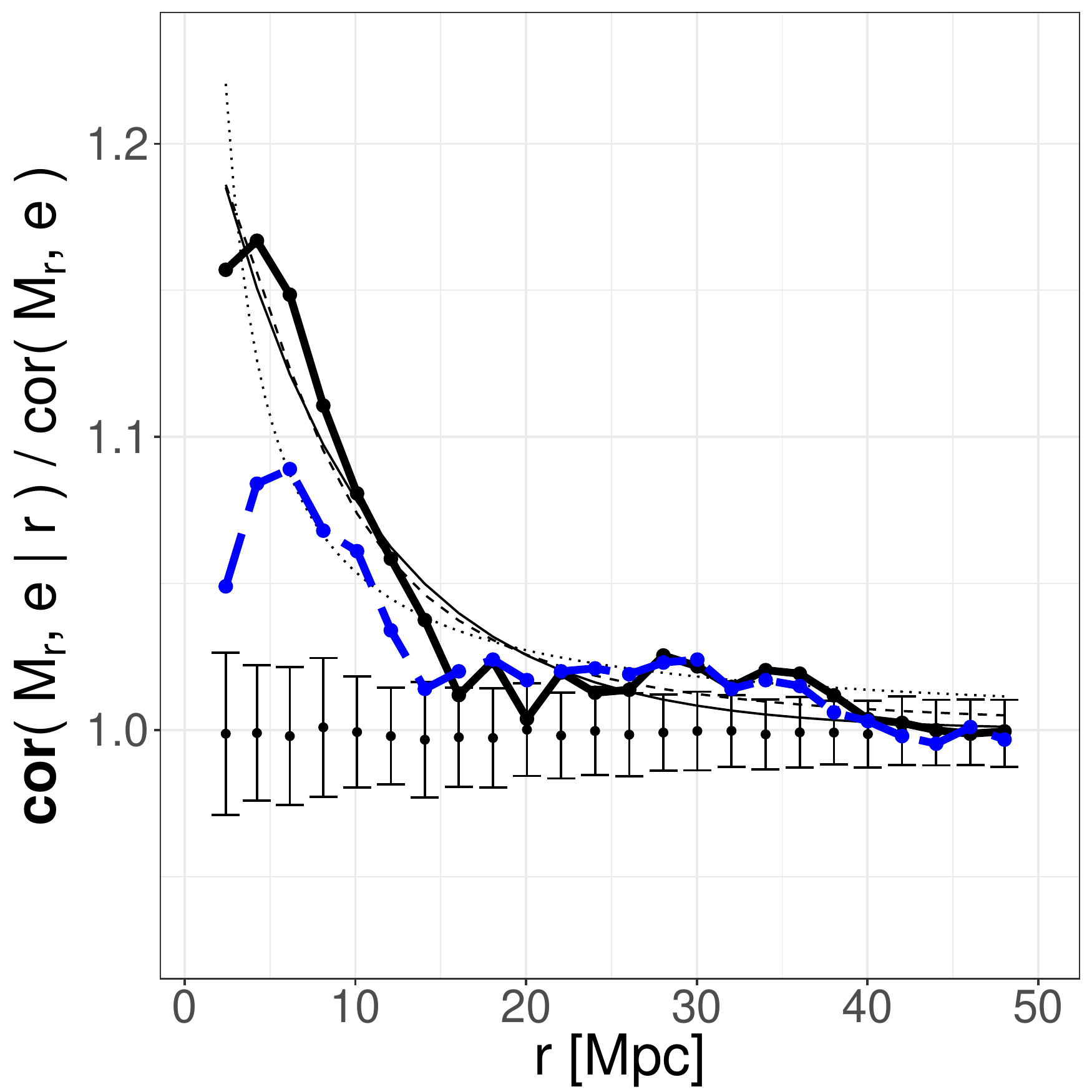}
  \includegraphics[width=5.5cm]{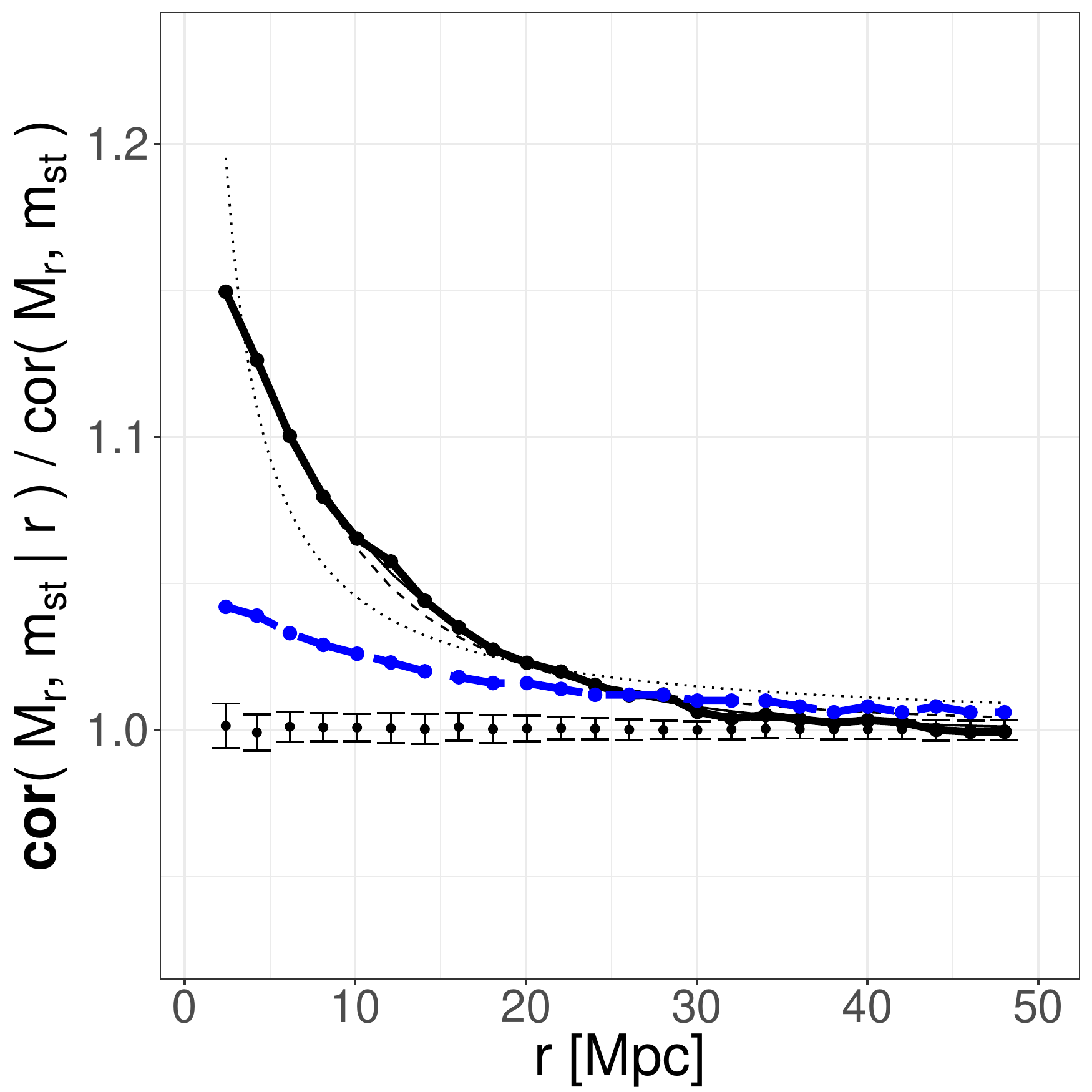}
  \includegraphics[width=5.5cm]{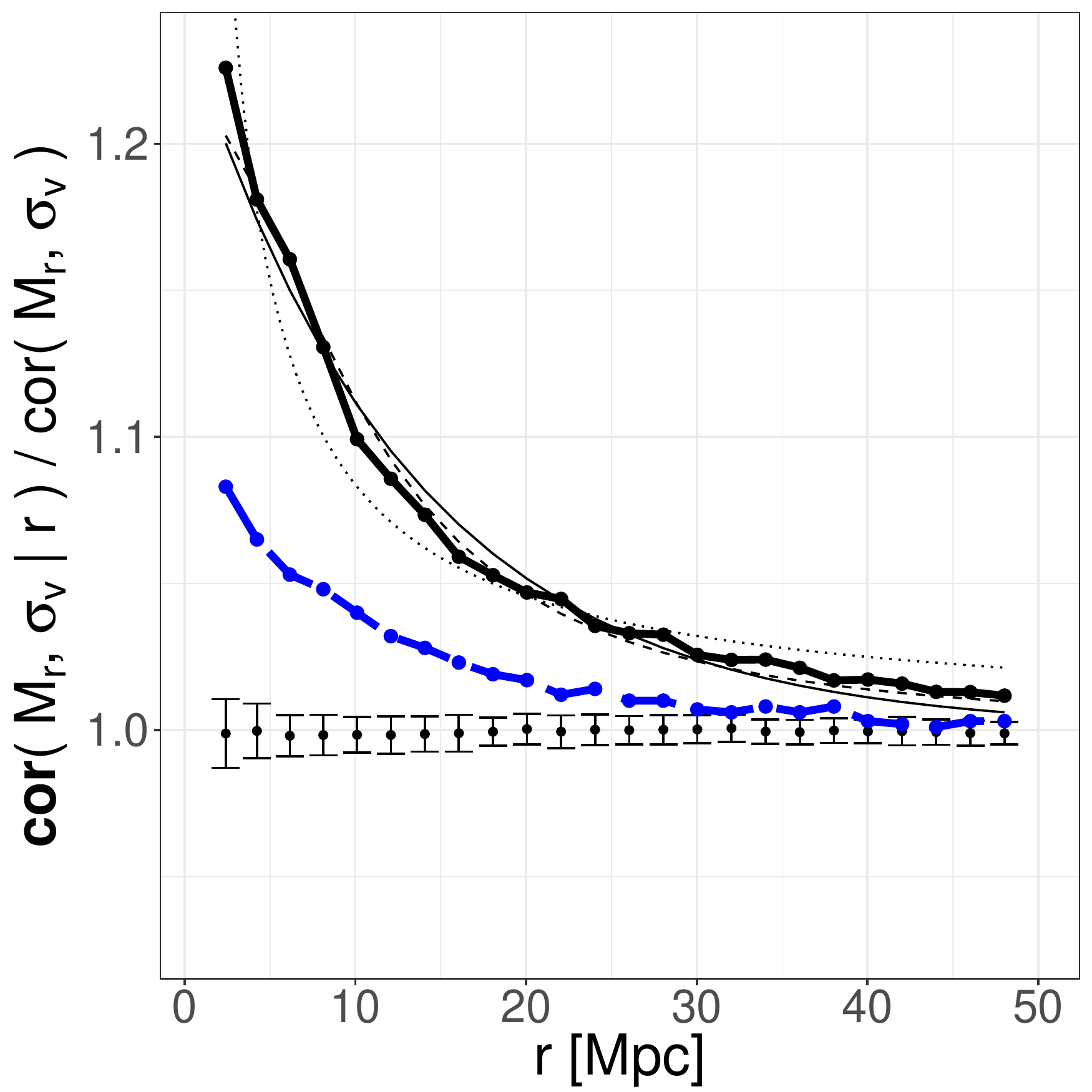}\\
  \includegraphics[width=5.5cm]{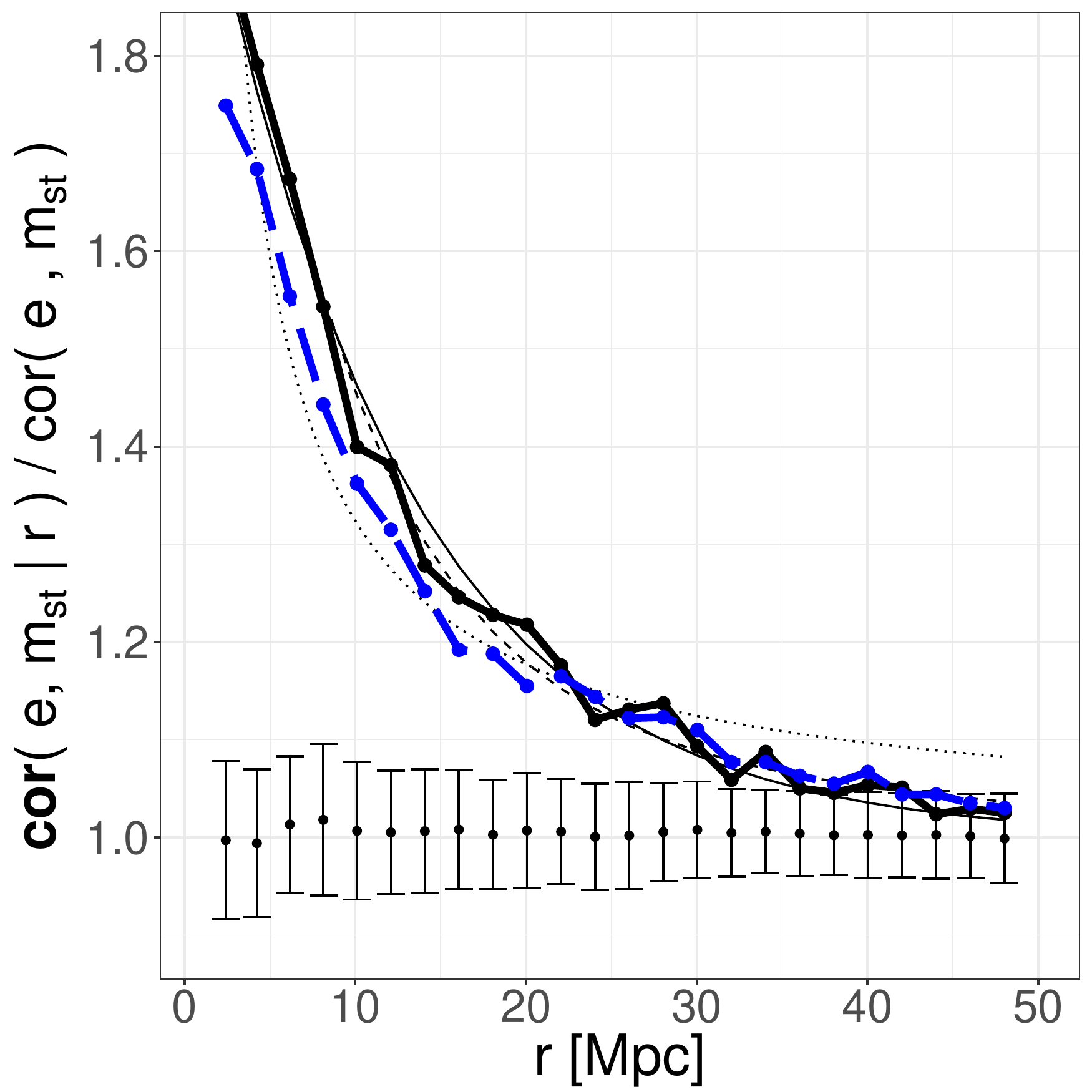}
  \includegraphics[width=5.5cm]{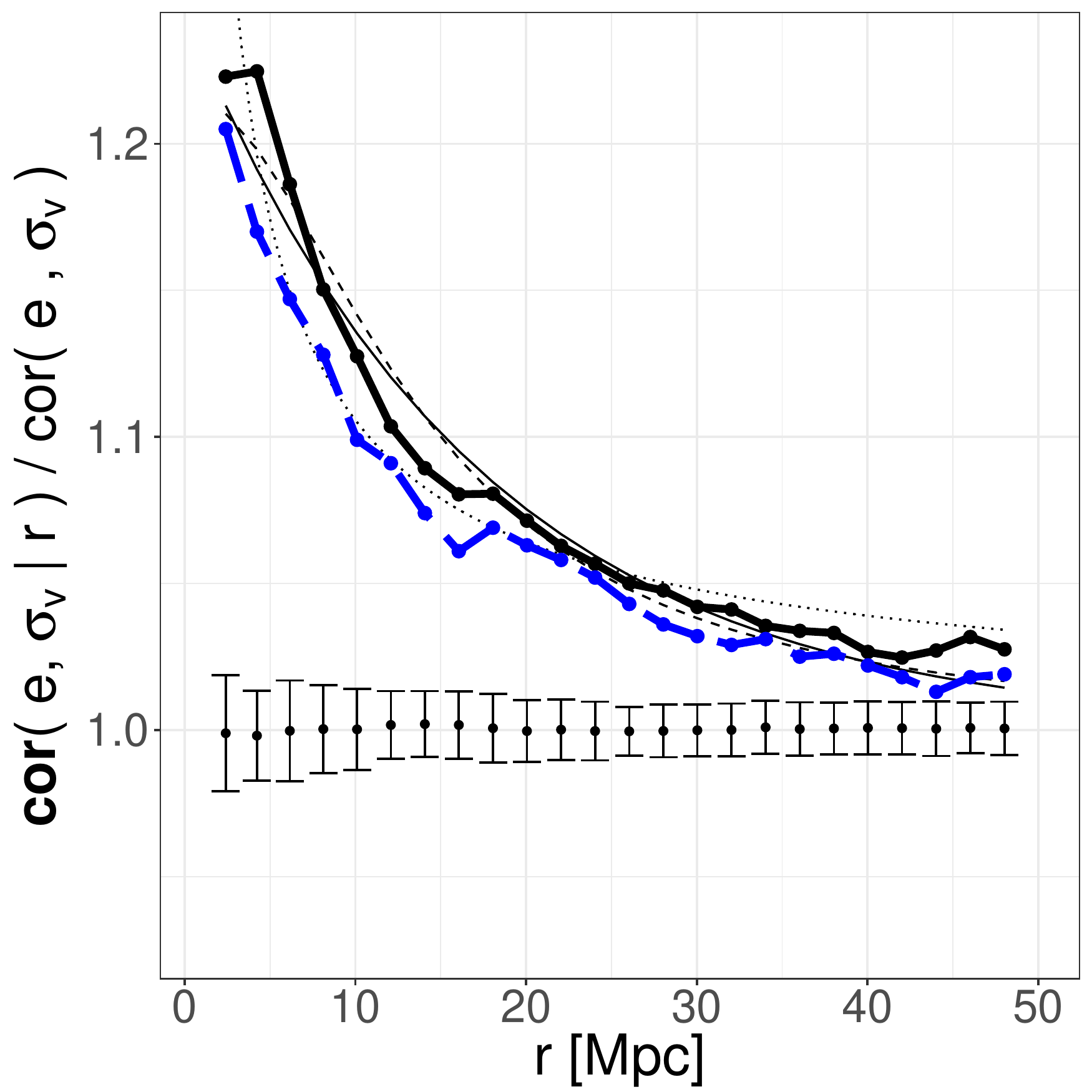}
  \includegraphics[width=5.5cm]{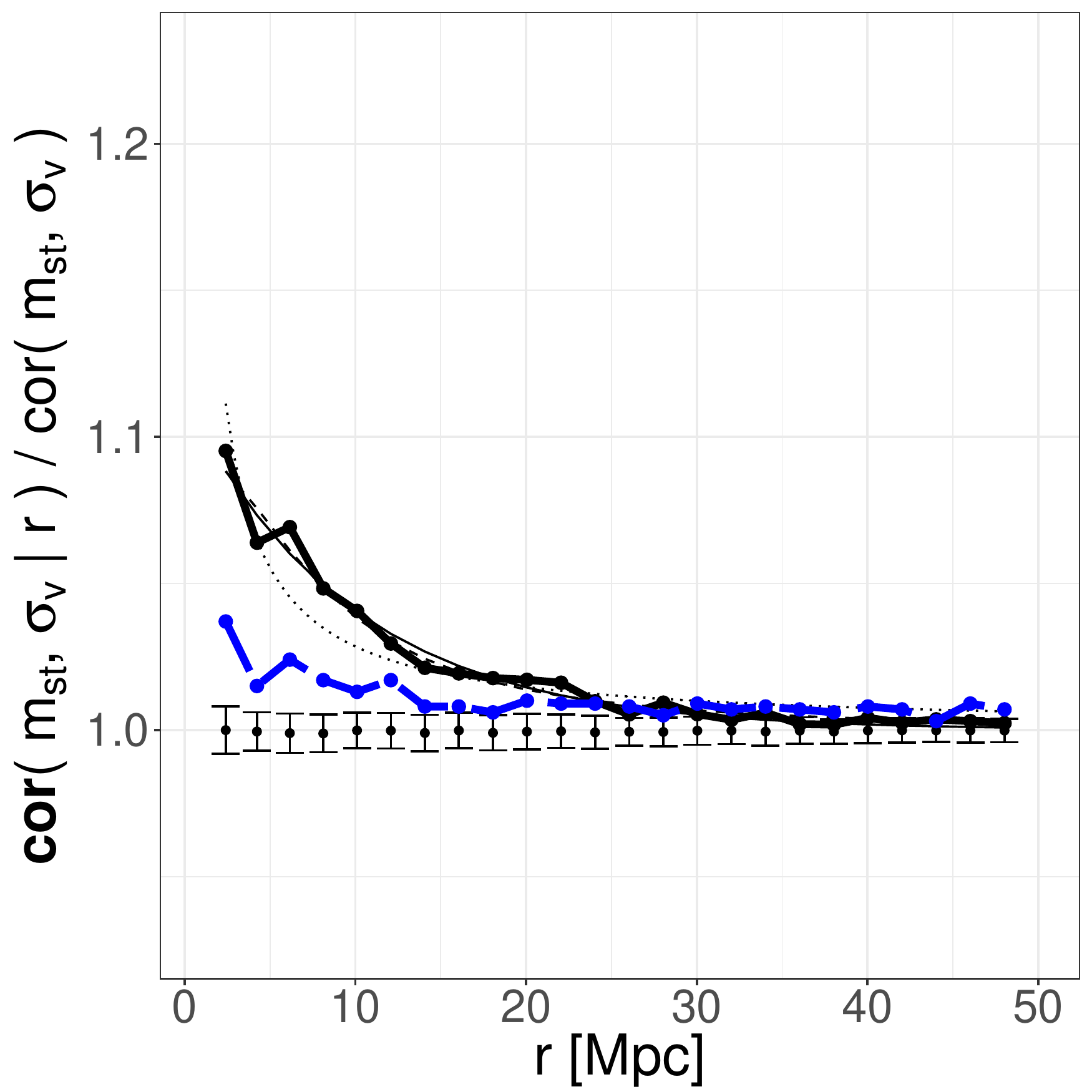}
  \caption{The scale dependent correlation functions 
    $\textbf{cor}(\cdot,\cdot|r)/\text{cor}(\cdot,\cdot)$ of $(M_r,e)$,
    $(M_r,m_\text{st})$, $(M_r,\sigma_v)$, $(e,m_\text{st})$,
    $(e,\sigma_v)$, and $(m_\text{st},\sigma_v)$ calculated from the
    volume limited sample with 600\,Mpc depth from the SDSS DR12 (thick
    solid line).
    The one--$sigma$ error bars around 
    $\textbf{cor}(\cdot,\cdot|r)=\text{cor}(\cdot,\cdot)$ are calculated 
    from 50~galaxy samples with randomised marks.
    The exponential, Lorentz and power--law fits, according to
    Eq.\,\ref{eq:fit}, are shown with a thin solid, dashed and dotted
    line respectively. 
    The thick blue dashed line shows the results for the alternative 
    definition of the scale dependent correlation coefficients  
    $\widetilde{\textbf{cor}}(\cdot,\cdot|r)/\text{cor}(\cdot,\cdot)$.
    The one--$\sigma$ error bars for
    $\widetilde{\textbf{cor}}(\cdot,\cdot|r)$ (not shown in the plot)
    show a very similar amplitude in comparison to the errors
    calculated for $\textbf{cor}(\cdot,\cdot|r)$.}
    \label{fig:sdssmarkcor}
\end{figure*}

\begin{table}
  \caption{The correlation coefficients (Eq.\,\ref{eq:cor-onepoint})) 
    between $M_r$, $e$, $\sigma_v$ and $m_\text{st}$ determined from 
    the volume limited sample with 600\,Mpc depth from the SDSS~DR12.}
  \label{tab:sdsscorrmatv600}
\centering
\begin{tabular}{c || c c c c}
   cor$(\cdot,\cdot)$  & $M_r$ & $e$ &  $m_\text{st}$ & $\sigma_v$\\
\hline\hline
$M_r$  & 1 & 0.15 & -0.84 & -0.49\\
$e$      &  & 1 & -0.05 & -0.2 \\
$m_\text{st}$  & & & 1 & 0.65 \\
$\sigma_v$  & & & & 1
\end{tabular}
\end{table}
The (one-point) correlation coefficients (Eq.\,(\ref{eq:cor-onepoint}))
between the marks $M_r$, $e$, $\sigma_v$ and $m_\text{st}$ in the
volume limited galaxy sample with 600\,Mpc depth are shown in
Table\,\ref{tab:sdsscorrmatv600}. These sometimes strong
(anti-)\,correlations are expected. E.g.\ the absolute magnitude $M_r$
is the negative logarithm of the luminosity, hence a strong
anti--correlation with the logarithmic stellar mass $m_\text{st}$ is
anticipated.
\begin{figure}
  \centering
  \includegraphics[width=4.35cm]{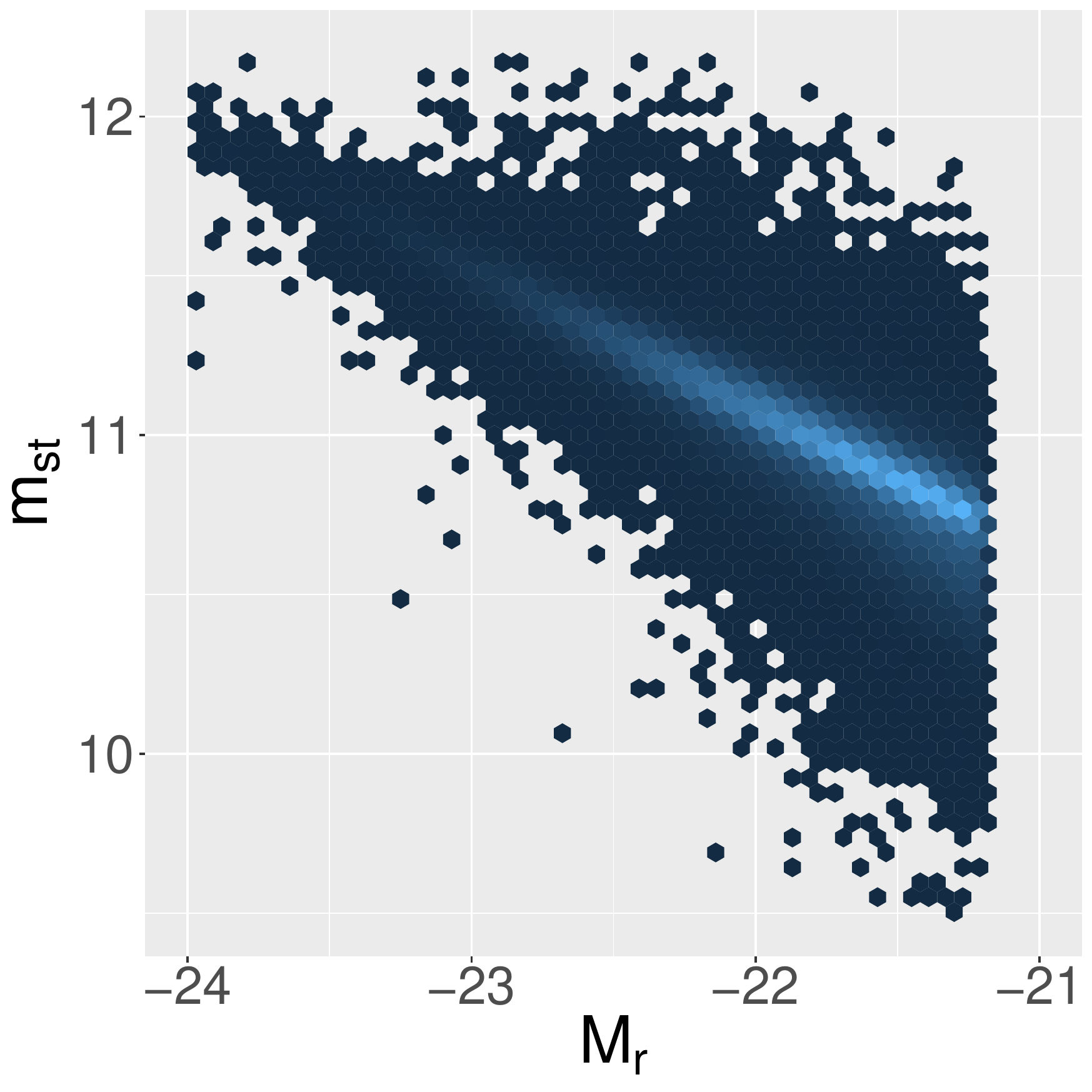}
  \includegraphics[width=4.35cm]{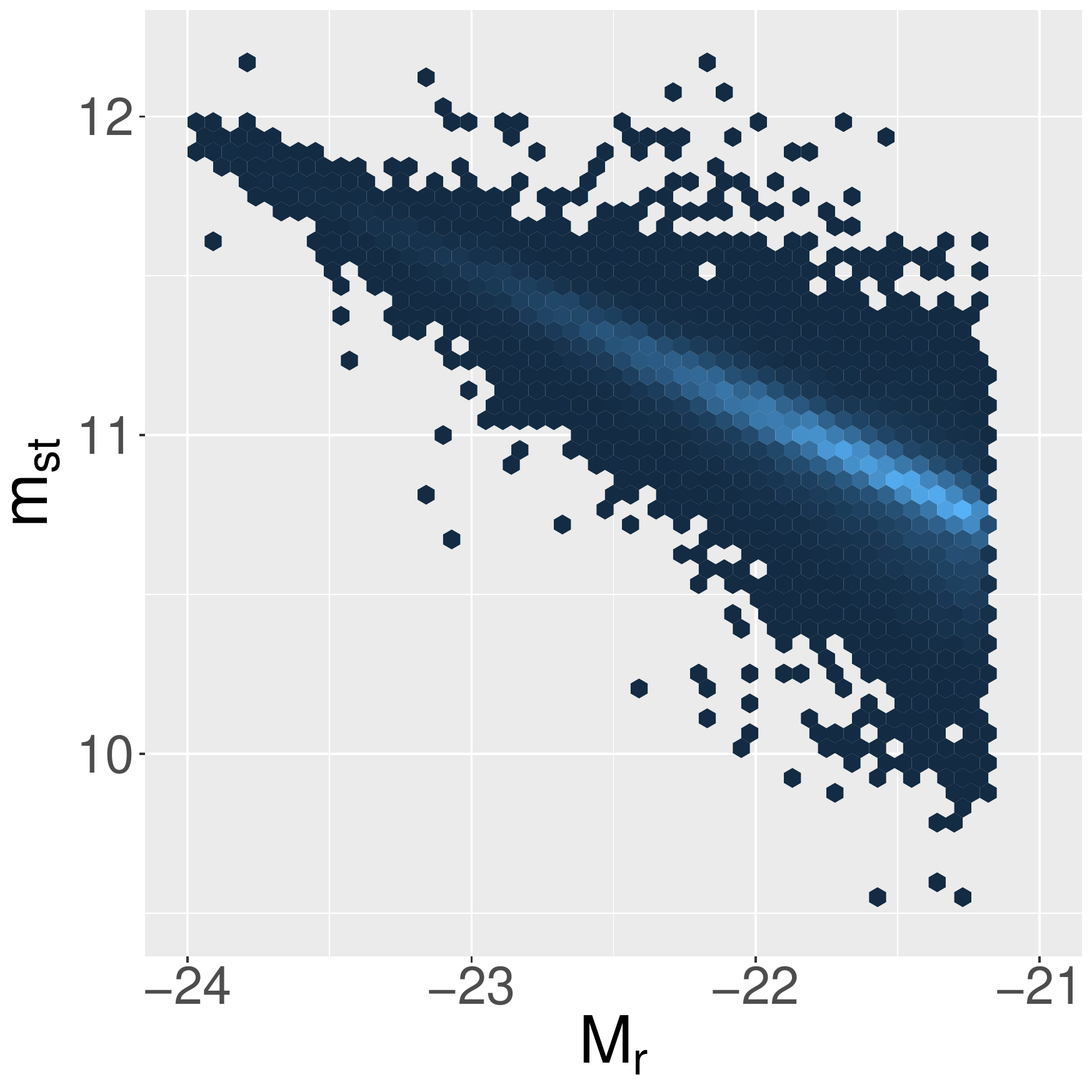}
  \caption{In both plots the relative frequencies of galaxies in the
    $(M_r, m_\text{st})$--plane are shown. The brighter the colour,
    the more galaxies are within this pixel. The left plot is with all
    galaxies, whereas the right plot only shows galaxies with a
    neighbouring galaxy at a distance of $r\in[1,3]$\,Mpc. The
    normalisation of the logarithmic quantities $M_r$ and
    $m_\text{st}$ is given in Appendix~\ref{sec:samples-sdss}}.
    \label{fig:sdsshists}
\end{figure}
This strong anti--correlation between $M_r$ and $m_\text{st}$ is also
clearly visible from the 2d--histogram in Fig.\,\ref{fig:sdsshists}.
Moreover, galaxies in close pairs show an even stronger
anti--correlation between $M_r$ and $m_\text{st}$, as seen from the
tighter histogram for the close pairs.  Exactly this visual impression is
quantified with the scale dependent correlation coefficient
$\textbf{cor}\left(M_r,m_\text{st}|r\right)$.
In Fig.\,\ref{fig:sdssmarkcor} the
$|\textbf{cor}\left(M_r,m_\text{st}|r\right)|\gg|\text{cor}(M_r,e)|$
shows the tightened correlation for close pairs (small $r$), whereas
the scale dependent correlation coefficient approaches the overall
average $\text{cor}(M_r,e)$ for large $r$.
This increased correlation of
$|\textbf{cor}\left(M_r,m_\text{st}|r\right)|$ compared to
$|\text{cor}(M_r,e)|$ is the scale dependent signal of galactic
conformity.

The scale dependent correlation coefficients are shown in
Fig.\,\ref{fig:sdssmarkcor} for the six combination of the marks
$M_r$, $e$, $\sigma_v$ and $m_\text{st}$.  In all cases the modulus of
the scale dependent correlation coefficient, e.g.\
$\textbf{cor}\left(M_r,e|r\right)$ is significantly larger than the
modulus of the overall correlation coefficient
$\text{cor}\left(M_r,e\right)$ on small scales. On larges scales
$\textbf{cor}\left(M_r,e|r\right) \rightarrow
\text{cor}\left(M_r,e\right)$ as expected.
Randomising the marks, but keeping the positions fixed, allows us to
quantify the fluctuations around the case of mark independent
clustering.  For smaller distances $r$, the scale dependent correlation
coefficients are well outside the fluctuations of the randomised
samples -- a clear signal of galactic conformity.
This signal extends out to large scales, becoming consistent with mark
independent clustering beyond 40\,Mpc --- a long range
of galactic conformity.

\subsection{Determining the range}

\begin{table}
  \caption{The scale parameter $q$ and $q_L$ in [Mpc], determined from
    the exponential-- and Lorentz--fit (Eq.\,(\ref{eq:fit})) to the
    scale dependent correlation coefficients as shown in
    Fig.\,\ref{fig:sdssmarkcor}.}
\label{tab:sdssfits}
\centering
\begin{tabular}{l c c c c c c}
       & $M_r, e$ & $M_r, m_\text{st}$ &  $M_r, \sigma_v$ 
       & $e, m_\text{st}$ & $e, \sigma_v$ & $m_\text{st}, \sigma_v$\\
\hline\hline
$q$    & 8.89 & 9.35 & 13.0 & 11.7 & 16.9 & 9.76\\
\hline
$q_L$  & 7.58 & 7.82 & 10.5 & 9.65 & 13.9 & 8.2
\end{tabular}
\end{table}
To quantify the range of conformity, an exponential, a Lorentz
function and a power--law are fitted\footnote{The weighted least-square
  fit is performed with the function \texttt{nls} from the statistic
  package \texttt{R} using the inverse variance of the randomised
  sample as weights.} to the observed scale dependent correlation
coefficients:
\begin{equation}
 \textbf{cor}(m_1,m_2|r)-1\ \sim\
\left\{\  A\exp\left(-\frac{r}{q}\right), \quad
A'\frac{q_L}{q_L^2 + r^2},\quad
A'' r^\gamma\ \right\}.
\label{eq:fit}
\end{equation}
$q$ and $q_L$ are the scale parameters in the exponential-- and
Lorentz--model, the power--law is scale invariant.
As can be seen from Fig.\,\ref{fig:sdssmarkcor} in all six cases the
exponential-- and the Lorentz--fit perform similarly well, whereas the
scale invariant power--law fit is significantly off. Quantitatively
this can be seen from the summed residuals. For the exponential-- and
Lorentz--fit they are comparable in size, wheres for the power-law fit
they are larger by an order of magnitude.
The $q$ and $q_L$ determined from fits are ranging from 8\,Mpc to
17\,Mpc (see Table\,\ref{tab:sdssfits}).  This quantifies the visual 
expression from Fig.\,\ref{fig:sdssmarkcor}, that the range of conformity 
depends on the galactic properties under investigation.
An exponential-- or a Lorentz--distribution function allows signals on
scales larger than $q$ and $q_l$. And indeed significant scale
dependent correlation coefficients are seen up to 40\,Mpc and beyond
(c.f.\ Fig.\,\ref{fig:sdssmarkcor}).
The toy model in Appendix~\ref{sec:toy} further illustrates that a
built in scale in the correlation pattern of the mark distribution can be
determined unambiguously with the scale dependent correlation
coefficients $\textbf{cor}(\cdot,\cdot|r)$.

\subsection{Alternative scale dependent correlation coefficients}

In Fig.\,\ref{fig:sdssmarkcor} also the results for the alternative
definition of the scale dependent correlation coefficients
$\widetilde{\textbf{cor}}(\cdot,\cdot|r)$ are shown.
The four combinations $(M_r,e)$, $(M_r,m_\text{st})$,
$(M_r,\sigma_v)$, and $(m_\text{st},\sigma_v)$ show a reduced
amplitude compared to $\textbf{cor}(\cdot,\cdot|r)$.
With $\widetilde{\textbf{cor}}(\cdot,\cdot|r)$ one is measuring the
scale dependent correlation coefficients with respect to the mean and
variance of the galaxies with another galaxy at a distance of $r$ (see
eqs.\,(\ref{eq:meanr}) and (\ref{eq:varr})).  With
$\textbf{cor}(\cdot,\cdot|r)$ the correlations are calculated with
respect to the mean and variance of all galaxies.
It is well known that for galaxies the scale dependent mean
$\overline{M_r}(r)$ and and variance $\sigma_{M_r}^2(r)$ are larger
than the overall mean $\overline{M_r}$ and variance $\sigma_{M_r}^2$
for small distances $r$ (see e.g.\ \citealt{beisbart:luminosity}).
Hence a reduced amplitude should be expected from
Eq.\,(\ref{eq:corrtilde}).  Still the remaining signal traced by
$\widetilde{\textbf{cor}}(\cdot,\cdot|r)$ shows a similar long range
of conformity outside the fluctuations.
Also the combinations $(e,m_\text{st})$ and $(e,\sigma_v)$ show no
significant deviation between
$\widetilde{\textbf{cor}}(\cdot,\cdot|r)$ and
$\textbf{cor}(\cdot,\cdot|r)$, both confirming the long range of
conformity.

\subsection{Systematics}
\label{sec:sdsssystematics}

\begin{figure}
  \centering
  \includegraphics[width=4.35cm]{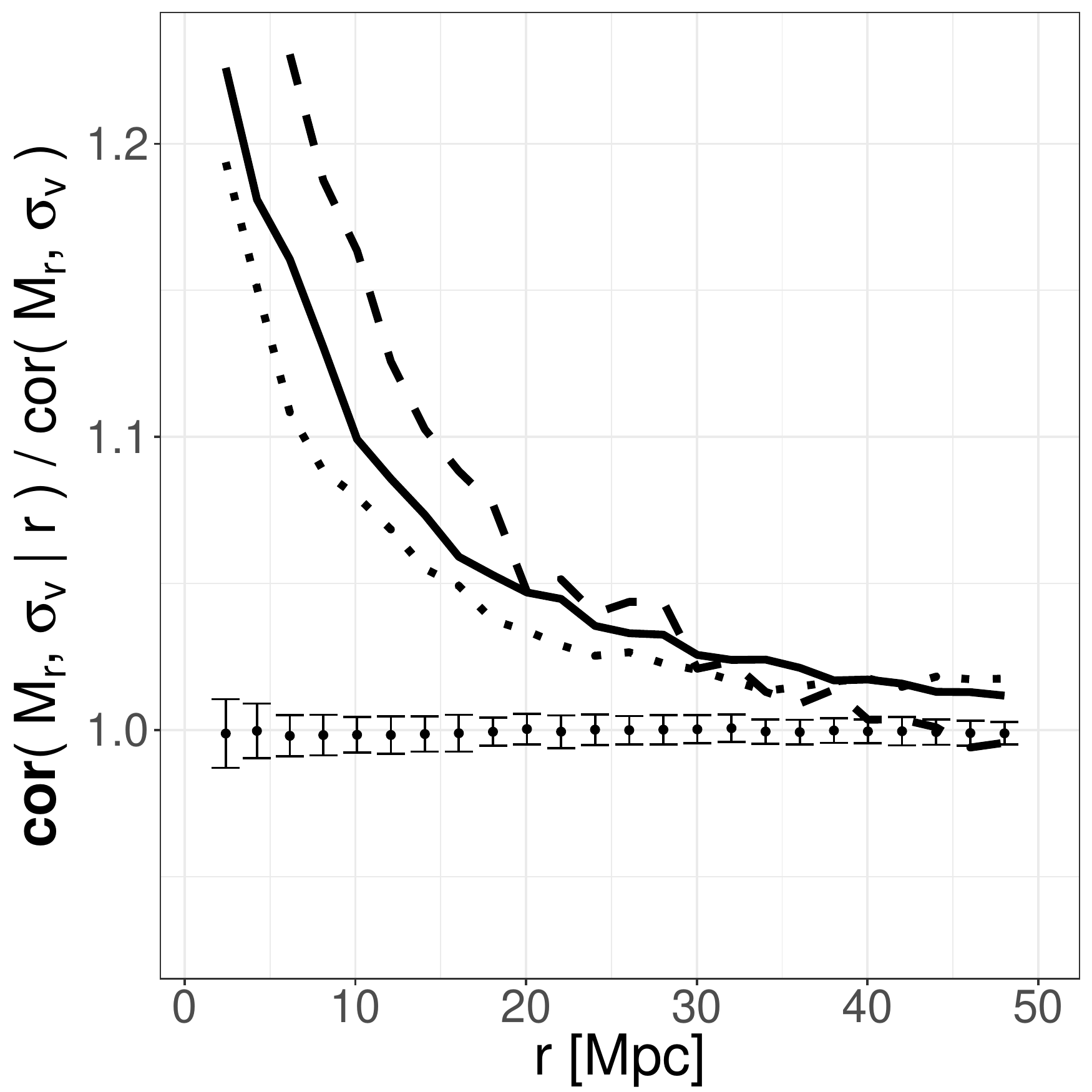}\hfill
  \includegraphics[width=4.35cm]{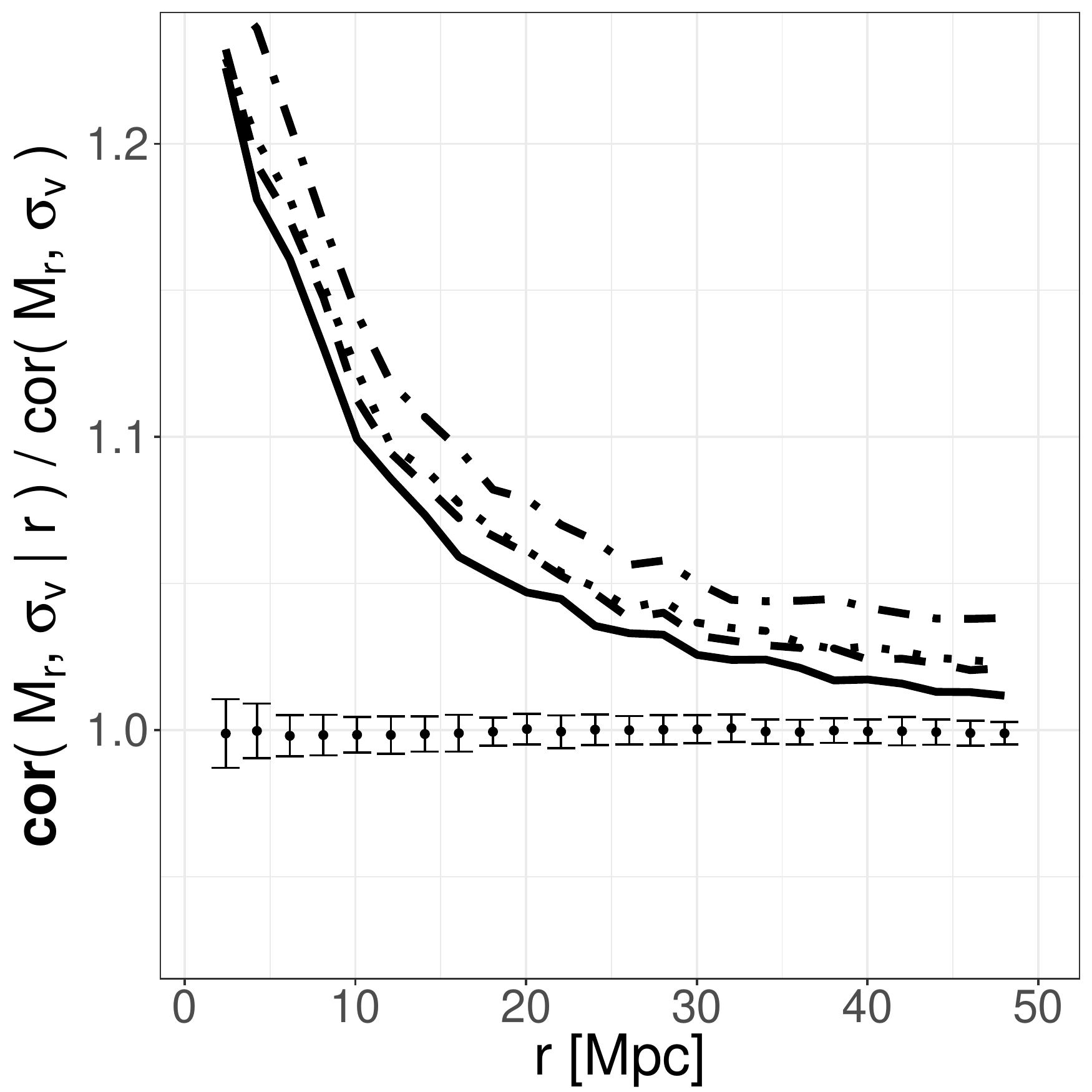}
  \caption{In the left plot $\textbf{cor}(M_r,\sigma_v\,|\,r)/\text{cor}(M_r,\sigma_v)$ 
    from volume limited
    samples with 300\,Mpc depth (dotted), 600\,Mpc depth (solid line), and
    900\,Mpc depth (dashed) are shown,
    in the right plot the results from the samples with different
    estimates of the absolute magnitude: K--corrected and extinction
    corrected (dereddened) model magnitudes (solid line), without
    extinction correction (dotted), without K--correction (dashed),
    without both, extinction correction and K--correction
    (dashed--dotted).}
  \label{fig:sdsssystematics}
\end{figure}

The results discussed in the preceding section, were obtained from a
volume limited galaxy sample from the SDSS~DR12 with a limiting depth
of 600\,Mpc. In Fig.\,\ref{fig:sdsssystematics} similar patterns can
be observed for the scale dependent correlation coefficients from
samples with 300\,Mpc and 900\,Mpc depth.
A more detailed look shows that the inclusion of less luminous
galaxies in the 300\,Mpc sample leads to a smaller amplitude of the
scale dependent correlation coefficient and also a smaller estimate
for the scale parameters, whereas an increased amplitude is observed
for the more luminous galaxies in the 900\,Mpc sample. The amplitude
and range of conformity is not universal, it depends on the galactic
properties considered and on the luminosity cut used for the
construction of the sample.

The absolute magnitude $M_r$ is used as a mark but also used in the
construction of the volume limited samples. Hence it is important to
investigate how systematic changes in the calculation of $M_r$
influence the results.  The analysis was repeated for absolute
magnitudes derived from the model magnitudes with \emph{no} extinction
correction (dereddening) and\,/\,or without employing a
$K$--correction. As can be seen from Fig.\,\ref{fig:sdsssystematics},
the results are very similar, only the results from samples with no
extinction correction \emph{and} no--K--correction show a
significantly enhanced amplitude and an even longer range of
conformity.
To check for a special kind of Malmquist--bias (see
\citealt{beisbart:luminosity}, section~4.5), the analysis was repeated
for galaxies with a distance up to 580\,Mpc, selected from the volume
limited sample with limiting depth of 600\,Mpc and no significant
deviations were seen.

\begin{figure}
  \centering
  \includegraphics[width=4.35cm]{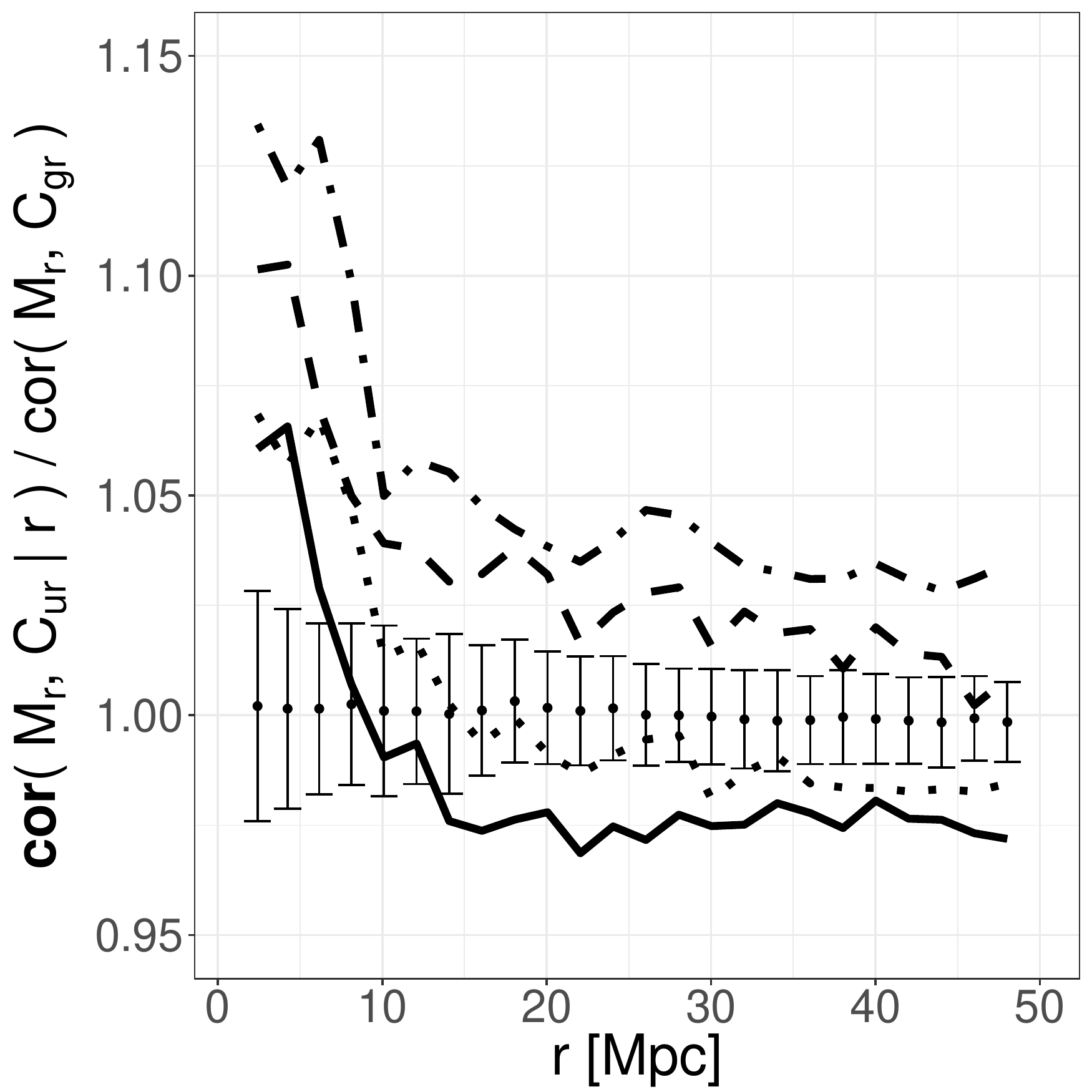}
  \caption{ The $\textbf{cor}(M_r,C_{ur}|r)/\text{cor}(M_r,C_{ur})$ as
    calculated from different estimates of the absolute Magnitude
    $M_r$ and colour $C_{ur}=M_u-M_r$: K--corrected and extinction
    corrected magnitudes (solid line), without extinction correction
    (dotted), without K--correction (dashed), without both, extinction
    correction and K--correction (dashed--dotted).}
  \label{fig:sdsscolours}
\end{figure}
The ratios of luminosities in different filters are called colours.
It is well known, that colours are correlated with the morphological
type and other properties of the galaxy. Hence colours should be
natural candidates in the analysis presented above.
However the scale dependent correlation coefficients for colours are
sensitive to the extinction correction and the K--correction.
Differences on small scales and residual correlations on large scale
can be seen for the colour $C_{ur}=M_u-M_r$ and the absolute magnitude
$M_r$ in Fig.\,\ref{fig:sdsscolours}.
The amplitude of the scale dependent correlation coefficient between
$M_r$, $e$, $\sigma_v$, and $m_\text{st}$, obtained from samples with
different magnitude estimates, differ slightly, but a consistent
picture for the conformity on large scales appears.  Hence, the main
results of this article, the long range of conformity, is not
affected as can be seen from Fig.\,\ref{fig:sdssmarkcor} and
Fig.\,\ref{fig:sdsssystematics}.  Moreover, the sample using the
extinction and K--corrected magnitudes gives the most conservative
estimates for the scale dependent correlation coefficients with the
lowest amplitude and the smallest range of conformity.
Unfortunately this is not the case for the scale dependent correlation
coefficients of colour $C_{ur}$ and absolute magnitudes $M_r$ (see
Fig.\,\ref{fig:sdsscolours}). It is not clear whether the extinction
correction, the K--correction, or other currently unknown issues are
responsible for these residuals and therefore colours are not
considered any further in this work.

Instead of colours the spectral properties of the galaxies can be used
directly. E.g.\ from the observed line--widths one estimates the
velocity dispersion $\sigma_v$ in a galaxies.
Also the stellar mass estimates rely heavily on spectral properties of
the galaxies, and one may think of the stellar mass estimate as a
concise summary of the spectral properties of the galaxy.
As briefly discussed in Appendix~\ref{sec:samples-sdss} different
methods employing different spectral libraries can be used to estimate
the stellar mass content $m_\text{st}$. Repeating the analysis for the
three different stellar mass estimates from the SDSS database leads to
very similar results.

\section{Scale dependent correlation coefficients of halos 
from the MultiDark simulations}
\label{sec:simulation}

Dark matter simulations can be used to model the large scale
distribution of matter in the universe. The dark matter
concentrations in these simulations are called halos.  A direct
comparison of the result for galaxies to the results from halos is
complicated by the fact that no luminous matter is included in the
simulations. Still, analog properties of the halos can be used
and the scale dependent correlation coefficients calculated from 
dark matter halos can be qualitatively compared to the results from 
the galaxies.  The focus is on the range of these scale dependent 
correlation coefficients.
A related motivation for investigating halo catalogs is coming from
the observations in the galaxy catalog that there are residuals in the
scale dependent correlation coefficients for colours which are not
well understood (see Sect.~\ref{sec:sdsssystematics}).  The scale
dependent correlation coefficients for the other galactic properties
do not show these residuals but still one wishes for an, at least
qualitative cross check.  Halo catalogs from dark matter simulations
offer such clean well defined samples without observational biases.

From the MultiDark Simulations (MDPL2,
\citealt{prada:haloconcentration,klypin:multidark}) dark matter halos
are identified using the Rockstar halo--finder
\citep{behroozi:rockstar}.  Halos with a virial mass
$M_\text{vir}\ge10^{12}M_\odot/h$ (hence with at least 662 dark matter
particles per halo) are selected from the MDPL2 simulations.
The Rockstart halo--finder is able to determine sub--halos within
halos.  However in this analysis only distinct halos, i.e.\ halos
which are not a sub--halo in any other halo are used.  See
\citet{behroozi:rockstar}, Sect.\,3.4 for a detailed description of
how the substructure membership is determined.
The virial mass $M_\text{vir}$ and the dimensionless spin parameter
$\lambda$ of the halos are used as marks, and the ratio of the
smallest axes to the largest axes in the mass ellipsoid (for details
see Appendix~\ref{sec:samples-multidark}).
No direct comparison of the scale dependent correlation coefficients
from the dark matter halos and the galaxy distribution is attempted,
but analog quantities are used as marks: For the dark matter
halos from the simulations the mass is directly accessible, whereas
for galaxies the absolute magnitude and the stellar mass content are
biased tracers of the overall mass.  The internal dynamical state is
reflected in the spin of the halo and in the velocity dispersion of
the galaxy.  The shape of the halo is quantified from the 3d--mass
ellipsoid, and the shape of a galaxy from the 2d--ellipticity obtained
from the image of the galaxy.

\begin{table}
  \caption{The correlation coefficients between $M_\text{vir}$,
    $\lambda$, and $s$ determined from the MDPL2 Rockstar 
    halo sample with $M_\text{vir}\ge10^{12}M_\odot/h$.}
  \label{tab:corrmatmdpl2}
\centering
\begin{tabular}{c || c c c c}
   cor$(\cdot,\cdot)$  & $M_\text{vir}$ & $\lambda$  & $s$\\
\hline\hline
$M_\text{vir}$  & 1 & -0.03 & -0.12\\
$\lambda$      &  & 1 & -0.30 \\
$s$  & & & 1
\end{tabular}
\end{table}
Table\,\ref{tab:corrmatmdpl2} summarises the correlation coefficients
between $M_\text{vir}$, $\lambda$, and $s$ in the halo sample.  Such
correlations are expected. For a detailed study of these one point
correlations see e.g.\ \citet{knebe:oncorrelation} and
\citet{vega-ferrero:shape}.
\begin{figure*}
  \centering
  \includegraphics[width=5.4cm]{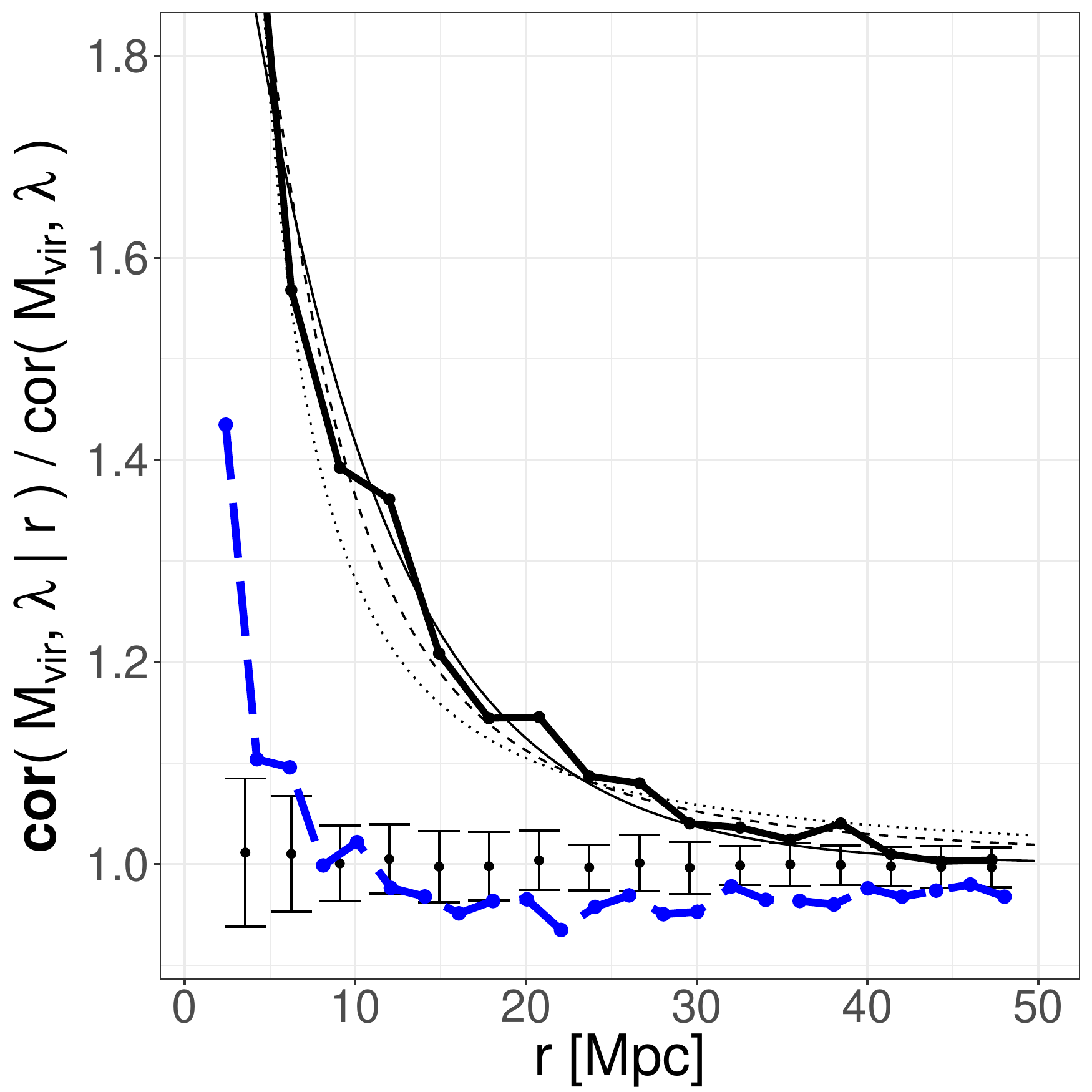}
  \includegraphics[width=5.5cm]{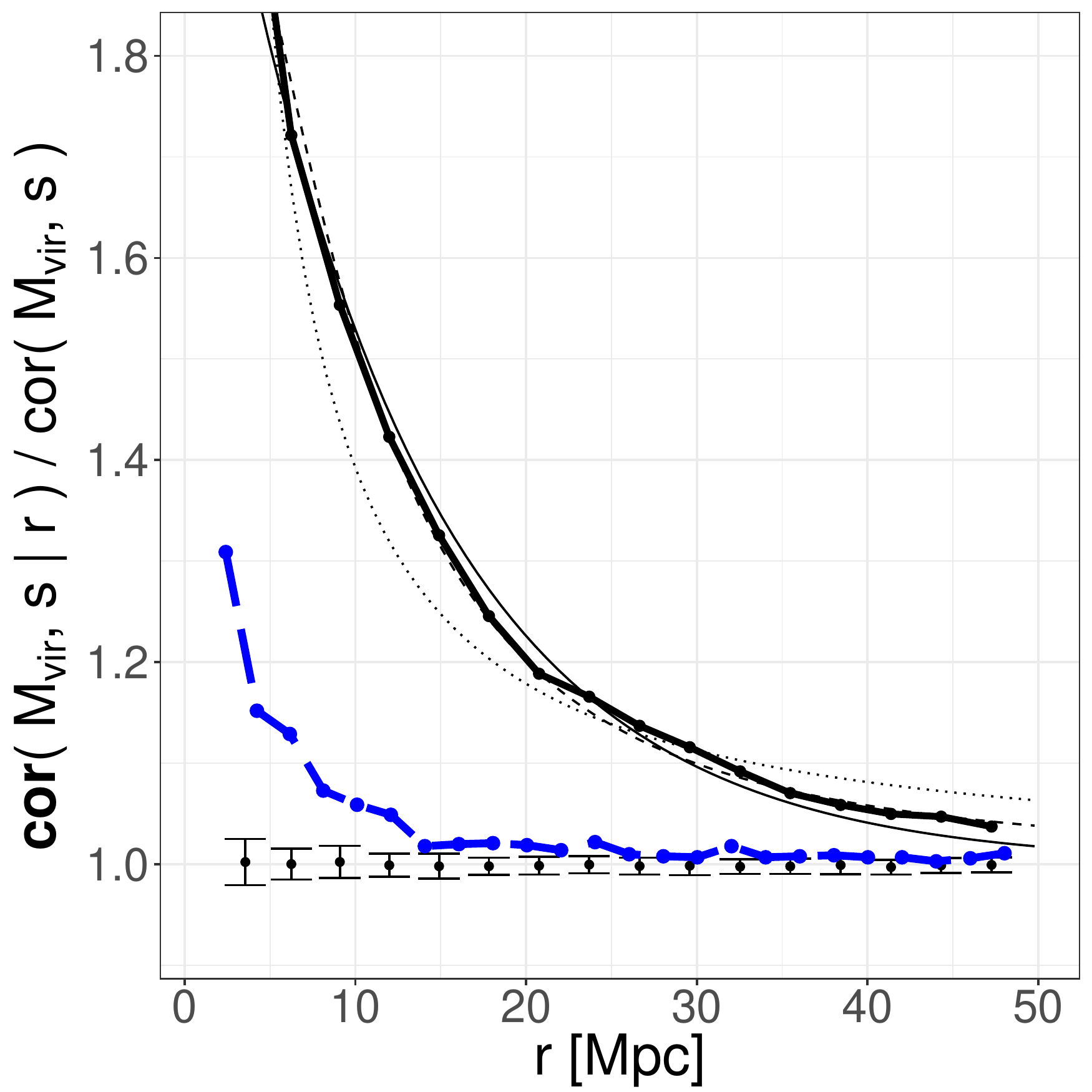}
  \includegraphics[width=5.5cm]{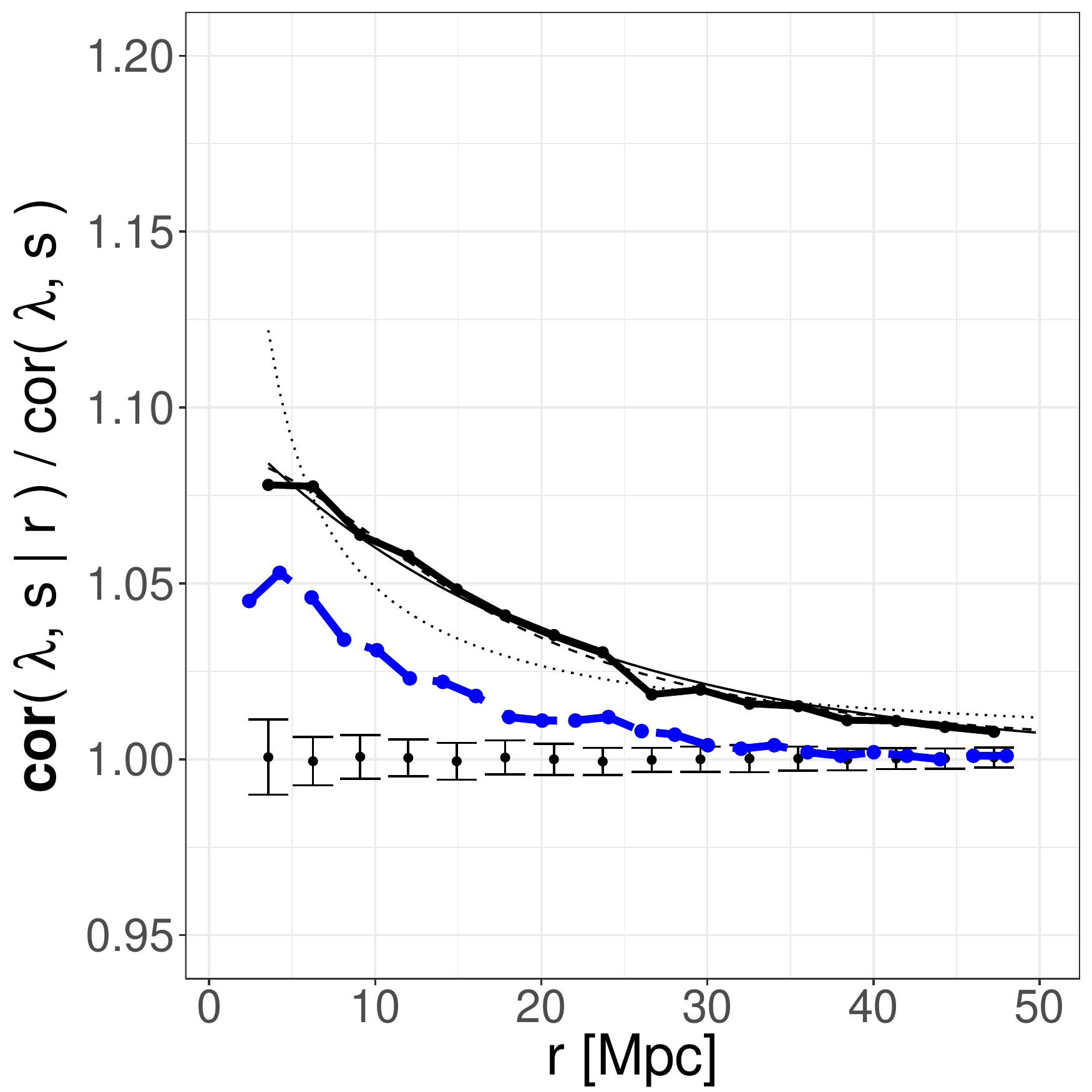}
  \caption{The scale dependent correlation functions of
    $(M_\text{vir}, \lambda), (M_\text{vir}, s), and (\lambda, s),$ calculated from the 
    MDPL2 Rockstar halo sample with $M_\text{vir}\ge10^{12}M_\odot/h$ (thick solid line).
    The one--$sigma$ error bars around 
    $\textbf{cor}(\cdot,\cdot|r)=\text{cor}(\cdot,\cdot)$ are calculated 
    from 50~halo samples with randomised marks.
    The exponential--, Lorentz-- and power-law fits, according to
    Eq.\,\ref{eq:fit}, are shown with thin solid, dashed and dotted
    lines respectively.}
    \label{fig:mdplmarkcor}
\end{figure*}
\begin{table}
  \caption{The scale parameter $q$ and $q_L$ in [Mpc], determined from the 
    exponential-- and Lorentz--fit (Eq.\,(\ref{eq:fit})) to the scale dependent 
    correlation coefficients as shown in Fig.\,\ref{fig:mdplmarkcor}.}
  \label{tab:mdplfits}
  \centering
  \begin{tabular}{l c c c}
         & $M_\text{vir}, \lambda$ &  $M_\text{vir}, s$ & $\lambda, s$ \\
\hline\hline
$q$      & 8.6 & 11.6 & 19.5 \\
\hline
$q_L$    & 6.1 & 9.0 & 16.8
\end{tabular}
\end{table}
Fig.\,\ref{fig:mdplmarkcor} shows the corresponding scale
dependent correlation coefficients.
The overall appearance is similar to the scale dependent correlation
coefficients observed in the galaxy distribution
(Fig.\,\ref{fig:sdssmarkcor}) with some exceptions.
The amplitude of the scale dependent correlation coefficients on small
scales is stronger for the combinations $(M_\text{vir},\lambda)$, and
$(M_\text{vir},s)$ compared to any of the results from the galaxy
distribution.
Also, the range of conformity is larger for the halos compared to the
galaxies --- see also the fitted scale parameters of the halo sample
in Table\,\ref{tab:mdplfits} compared to the scale parameters of the
galaxy sample in Table\,\ref{tab:sdssfits}.
Similar to the galaxy distribution, the alternative scale dependent
correlation coefficients $\widetilde{\textbf{cor}}(\cdot,\cdot|r)$
show a reduced amplitude. Still $(\lambda,s)$ shows long
range correlations out to 30\,Mpc, but the signal in
$(M_\text{vir},\lambda)$ and $(M_\text{vir},s)$ is confined to scales
below 10~and 15\,Mpc.

\subsection{Systematics}

\begin{figure}
  \centering
  \includegraphics[width=4.35cm]{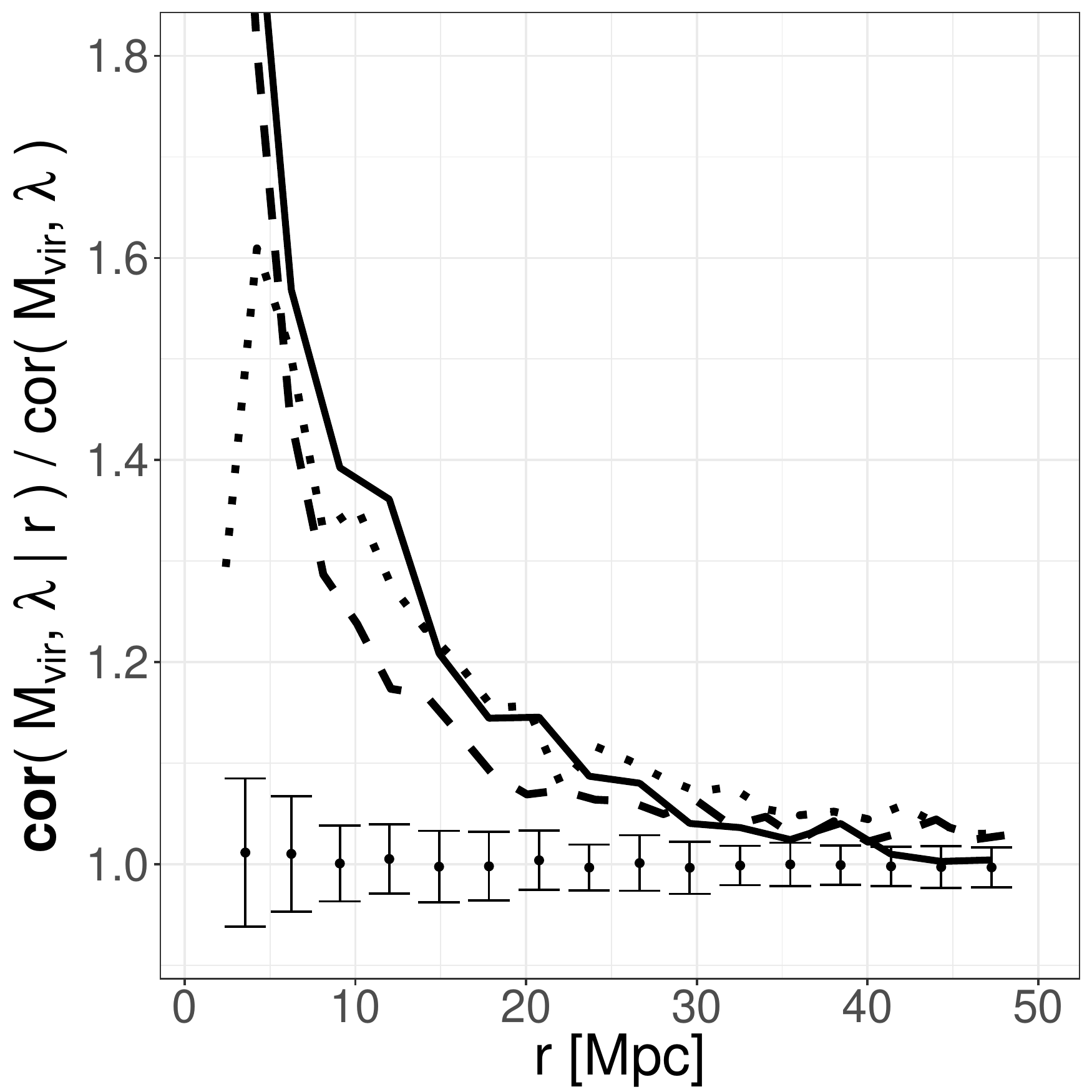}
  \hfill
  \includegraphics[width=4.35cm]{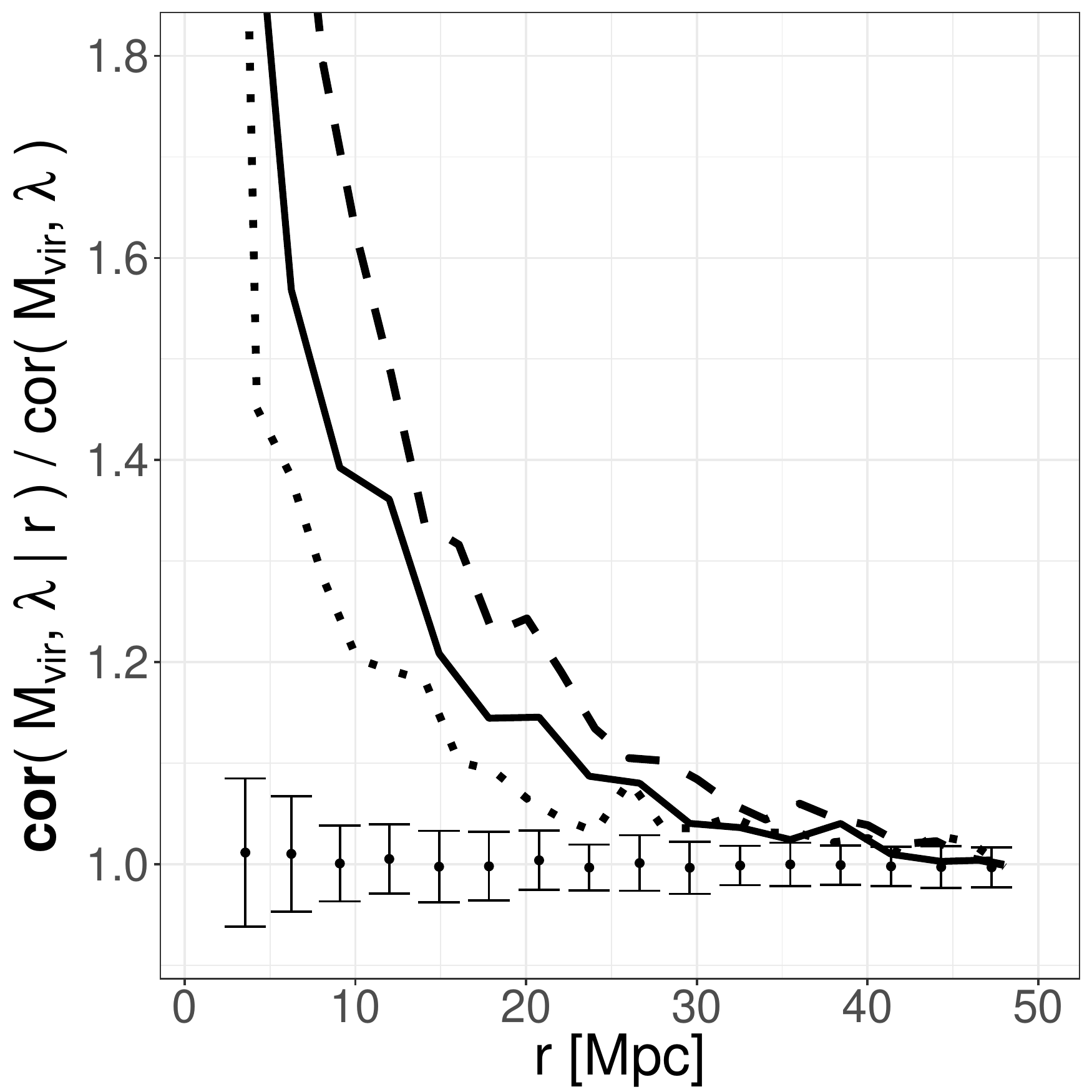}
  \caption{ In the left plot
    $\textbf{cor}(M_\text{vir},\lambda\,|\,r)/\text{cor}(M_\text{vir},\lambda)$ 
    are shown for samples with a mass--cut 
    $M_\text{vir}\ge5\times10^{11}M_\odot/h$ (dashed), 
    $M_\text{vir}\ge1\times10^{12}M_\odot/h$ (solid line), and 
    $M_\text{vir}\ge10^{13}M_\odot/h$ (dotted).
    In the right plot the results from samples using different halo
    identification methods are shown: rockstar distinct halos (solid
    line), rockstar all halos (dotted), FoF halos (dashed).}
  \label{fig:mdplsystematics}
\end{figure}

To investigate the dependence on the mass-cut, samples with
$M_\text{vir}\ge5\times10^{11}M_\odot/h$,
$M_\text{vir}\ge1\times10^{12}M_\odot/h$, and
$M_\text{vir}\ge10^{13}M_\odot/h$ have been analysed.  The scale
dependent correlation coefficients show a similar shape and in most
cases a similar amplitude between the halo sample.
As can be seen in Fig.\,\ref{fig:mdplsystematics} the amplitude and
range of conformity is increasing in the two samples with the mass cut
from $M_\text{vir}\ge5\times10^{11}M_\odot/h$ to
$M_\text{vir}\ge1\times10^{12}M_\odot/h$. A similar behaviour can be
observed in the galaxy samples including more luminous galaxies (see
Fig.\,\ref{fig:sdsssystematics}).
The most massive sample with $M_\text{vir}\ge1\times10^{13}M_\odot/h$
shows a dip in the scale dependent correlation coefficient on scales
below 5\,Mpc but very similar results compared to the sample with
$M_\text{vir}\ge1\times10^{12}M_\odot/h$ on large scale.
Also the scale dependent correlation coefficients of halo samples from
the BigMDPL simulations (box--size 2.5~Gpc/h) show a similar long
range of conformity.

The Rockstar halo--finder is able to determine a halo hierarchy.  In
the analysis for Fig.\,\ref{fig:mdplmarkcor} only distinct halos,
i.e.\ halos which are not marked as a sub--halos, are used. The scale
dependent correlation coefficients calculated from \emph{all} the
halos, including sub--halos and their parent halos, show a reduced
amplitude as can be seen in Fig.\,\ref{fig:mdplsystematics}.
The Rockstar halo--finder uses phase-space information and an
elaborate unbinding strategy to define the halos. The
three--dimensional friend--of--friend (FoF) halo--finder operates only
in position space to identify halos as linked particle over--densities
\citep{riebe:multidark}.  The analysis with the scale dependent
correlation coefficients is repeated for such FoF halo samples from
the same MDPL2 simulation. Again the mass, the spin, and the axes
ratios of the ellipsoidal shape are used as marks (see
\citealt{riebe:multidark} for details). By comparing the corresponding
scale dependent correlation coefficient of Rockstar and FoF halo
samples, an increased amplitude can be seen in
Fig.\,\ref{fig:mdplsystematics}.
Although the amplitude of the scale dependent correlation coefficients
differ between ``all halos'', ``distinct halos'', and ``FoF-halos'',
the signal of a long range of conformity is clearly visible in all
the samples.

\section{Summary and Outlook}
\label{sec:summary}

Properties of galaxies show scale dependent correlation coefficients
out to large scales. Properties like mass and luminosity are
significantly stronger (anti-)\,correlated for close pairs compared to
the correlation coefficients in the overall sample. A clear signal of
conformity.
The analysis was carried out with a new descriptive statistic, the
scale dependent correlation coefficients.  They quantify how the
correlation coefficients between galactic properties vary under the
condition that another galaxy (or halo) is at a distance of $r$.
This signal of galactic conformity extends to large scales, in several
cases becoming consistent with mark independent clustering only beyond
40\,Mpc.  Several tests for systematic effects confirm the long range
of conformity.
Halo samples from dark matter simulations show a larger amplitude and
an even longer range of conformity. The scale dependent correlation
coefficients between e.g. mass and shape clearly deviates from the
overall correlation coefficient beyond 40\,Mpc.
No universal range of conformity is found. The range varies for
different properties under investigation and also depends on the
luminosity-- and mass--cut used in the construction of the samples.
Such a long range of conformity goes well with the investigations of
\cite{faltenbacher:halos}, who found alignment correlations for
cluster sized halos out to separations of 100\,Mpc/h.
The focus of the present investigation was on the introduction of the
scale dependent correlation coefficients and on the detection of a
long range of conformity. On small scales more complicated patterns
are expected and further investigations of the scale dependent
conformity should be accompanied by a detailed modelling.

Pure dark matter simulations capture only the gravitational part but
allow for a large number of halos and convincing statistics.  As shown
by \cite{gottloeber:shape} and \cite{teklu:connecting} there exists a
complex interplay between spin, mass and morphology of the dark matter
and the gas component within halos.  It will be highly interesting to
investigate the environmental dependence of such halos using the scale
dependent correlation coefficients.

Empirical relations, like the Tully--Fisher or the fundamental plane
relation are special correlations between the properties of a galaxy
(see e.g.\ \citealt{kelson:hstkeyXXVII, saulder:calibrating} and
references therein).  These empirical relations, like the fundamental
plane, depend on the amount of substructure in the objects (see
\citealt{fritsch:fp} for galaxy clusters). Hence one can expect that
an extended version of the scale dependent correlation coefficients
could be used to investigate the spatial scale dependence of such
empirical relations.

As already mentioned a detailed modelling of this signal of conformity
is the next step. Purely geometric models, like the toy--model in
Appendix~\ref{sec:toy} help us to appreciate the method, but often do
not promote a physical understanding. Hence clearly more physically
motivated models are needed.

Inspired by the ideas of hierarchical structure formation in dark
matter models the halo model was designed to explain the clustering of
galaxies (see \citealt{cooray:halo} for a review). The halo model is
able to reproduce the signal from the mark weighted correlation
function out to 20\,Mpc (\citealt{skibba:luminosity}, see also
\citealt{paranjabe:correlating, pahwa:analytical} for a more detailed
model of galactic conformity). Within these models the contribution
from the so-called 2--halo term seems necessary to explain conformity
on large scales.
A physical explanation of galactic conformity from structure formation
is given by \cite{hearin:physicalorigin}, also called assembly bias.
Their explanation is elaborated for pair distances below 10\,Mpc, but
possibly their arguments could be extended to larges scales, too.

Another approach is based on the peak theory
\citep{bardeen:gauss}. Recently \citet{verde:bias} calculated the
Lagrangian (formation) bias for a Gaussian density field.  The matter
density field can be approximated more reliable using a logarithmic
transformation \citep{falck:straigthening} which could serve as an
improved starting point for such a bias calculation.  Closely related
to the lognormal density field, the log--normal model for the galaxy
distribution \citep{coles:lognormal, moller:log} can be used as a
stochastic model for the point \emph{and} mark distribution. For such
an intensity marked point process, the mark correlation functions can
be calculated explicitly \citep{ho:modelling,
  myllymaeki:conditionally}. The adaption to the galaxy distribution
will reveal whether a natural parametrisation is possible within this
model.

\begin{acknowledgements}
  I would like to thank Stefan Gottl{\"o}ber for the hospitality and
  the discussions at the AIP.  I am grateful to Kristin Riebe and Ben
  Hoyle for support and information on using the CosmoSim-- and the
  SDSS--database respectively.
  Special thanks to Claus Beisbart and Alex Szalay, some of the ideas
  for this analysis emerged from discussions now more than ten years
  ago.
  For comments on the manuscript I would like to thank
  Thomas Buchert, Stefan Gottl{\"o}ber and Volker M{\"u}ller.
  I appreciate very much comments by Simon White, who suggested the
  alternative definition of the scale dependent correlation
  coefficient in Eq.\,(\ref{eq:corrtilde}) to me. 
  I would like to thank the anonymous referee for his constructive and 
  helpful comments.\\

  Funding for SDSS-III has been provided by the Alfred P. Sloan
  Foundation, the Participating Institutions, the National Science
  Foundation, and the U.S. Department of Energy Office of Science. The
  SDSS-III web site is http://www.sdss3.org/.

  SDSS-III is managed by the Astrophysical Research Consortium for the
  Participating Institutions of the SDSS-III Collaboration including the
  University of Arizona, the Brazilian Participation Group, Brookhaven
  National Laboratory, Carnegie Mellon University, University of
  Florida, the French Participation Group, the German Participation
  Group, Harvard University, the Instituto de Astrofisica de Canarias,
  the Michigan State/Notre Dame/JINA Participation Group, Johns Hopkins
  University, Lawrence Berkeley National Laboratory, Max Planck
  Institute for Astrophysics, Max Planck Institute for Extraterrestrial
  Physics, New Mexico State University, New York University, Ohio State
  University, Pennsylvania State University, University of Portsmouth,
  Princeton University, the Spanish Participation Group, University of
  Tokyo, University of Utah, Vanderbilt University, University of
  Virginia, University of Washington, and Yale University.

  This research made use of the ``K-corrections calculator'' service,
  especially the python code, available at http://kcor.sai.msu.ru/.

  The CosmoSim database used in this paper is a service by the
  Leibniz-Institute for Astrophysics Potsdam (AIP).  The MultiDark
  database was developed in cooperation with the Spanish MultiDark
  Consolider Project CSD2009-00064.

  The Bolshoi and MultiDark simulations have been performed within the
  Bolshoi project of the University of California High-Performance
  AstroComputing Center (UC-HiPACC) and were run at the NASA Ames
  Research Center. The Multidark Planck (MDPL) and the BigMD simulation
  suite have been performed in the Supermuc supercomputer at LRZ using
  time granted by PRACE.
 
  In the numerical analysis Python with scipy and for the plotting R
  with ggplot2 have been used \citep{jones:scipy,r:lnguage,hadley:ggplot2}.
\end{acknowledgements}

\bibliographystyle{aa}
\bibliography{my}

\appendix 

\section{Samples}
\label{sec:samples}

\subsection{Galaxay catalogues from the SDSS III, DR12}
\label{sec:samples-sdss}

\begin{figure}
\tiny
\begin{verbatim}
select s.ObjID, s.ra, s.dec, s.z, 
  dbo.fCosmoDl(s.z, 0.307115, 0.692885, 0.0, -1.0, 0.6777) 
       as lumdist,
  s.dered_u, s.dered_g, s.dered_r, s.dered_i, s.dered_z,
  gal.mE1_r, gal.mE2_r, so.velDisp,
  sm1.mstellar_median, sm2.mstellar_median, sm3.logMass
into mydb.maingal_properties
from dr12.SpecPhoto as s
  join dr12.Galaxy as gal on s.ObjID = gal.ObjID
  join dr12.SpecObj as so on s.specObjID = so.specObjID
  join dr12.stellarMassPCAWiscM11 as sm1 
       on s.specObjID = sm1.specObjID
  join dr12.stellarMassPCAWiscBC03 as sm2 
       on s.specObjID = sm2.specObjID 
  join dr12.stellarMassStarformingPort as sm3 
       on s.specObjID = sm3.specObjID
where
  s.type = 3 and s.class = 'galaxy' and s.zWarning = 0 and 
  s.petroMag_r <= 17.77
\end{verbatim}
\normalsize
\caption{The SQL code used on CasJobs to extract the basic sample from SDSS
  DR12.\label{fig:sqlsdss} }
\end{figure}

In the SDSS DR12 data release \citep{alam:dr12, eisenstein:sdssIII} each
galaxy comes with a wealth of properties.
The galaxy samples for the analysis are built in in two stages.
First, a basic galaxy sample is obtained from the SDSS database, then
derived quantities are calculated and the volume limited samples are
constructed.
Our basic galaxy sample was extracted from the SDSS database, as
provided via CasJobs: \url{http://skyserver.sdss.org/CasJobs/}, using
the SQL script shown in Fig.\,\ref{fig:sqlsdss}.  The query starts
with the view \texttt{SpecPhoto} and joins it with \texttt{Galaxy} and
\texttt{SpecObj} to gain access to further photometric and
spectroscopic parameters. The joins with the tables
\texttt{stellarMassPCAWiscM11~/ PCAWiscBC03~/
  stellarMassStarformingPort} are used to obtain the stellar mass
estimates. In the joins with the stellar mass tables some galaxies
could not be matched and 0.11\% of the galaxies are lost.
The function \texttt{fCosmoDl} provided in the SDSS database is used
to calculate the luminosity distance from the redshift, using a
Planck--like cosmology consistent with the MultiDark simulations, see
appendix~\ref{sec:samples-multidark}.
The selection in the \texttt{where} clause is mostly the original
selection as used for the SDSS main galaxy sample
\citep{strauss:spectroscopic}. 
From this basic sample the following parameters are calculated for
each galaxy. 
\begin{description}
\item{\textbf{Absolute magnitudes:}} The absolute magnitude $M_r$ in
  the $r$--band is calculated from the extinction corrected
  (dereddened) model magnitude $m_{r}$ using $M_r=m_r-D$, with the
  distance module $D=5 \log_{10}(d/10\text{pc})$ and the luminosity
  distance $d$ in pc.
  The absolute magnitude is $K$--corrected using the python code from
  \url{http://kcor.sai.msu.ru/}, version 2012, implementing the
  methods described in
  \cite{chillingarian:analytical,chillingarian:universal}.  
  See also the comparison of several
  $K$--corrections in \cite{omill:photometric}
\item{\textbf{Ellipticities:}} The ellipticities $e$ of the galaxies
  are calculated from the Stokes parameters $Q$ and $U$ using
  $e=1-\frac{1-Q^2U^2}{1+Q^2U^2}$. The Stokes parameters $Q$ and $U$
  have been estimated from the intensity profile of the galaxies in
  the $r$--band using the adaptive moments $mE1_r$ and $mE2_r$
  respectively.  \citep{bernstein:shapes}.  This ellipticity $e$ is an
  estimate of the observed 2D--ellipticity on the sky.  No attempt is
  made to derive a 3D/de--projected ellipticity.
\item{\textbf{Stellar mass content:}} Using the photometry and the
  spectra one can estimate the stellar mass content  of a
  galaxy.  The following three mass estimates can be retrieved from
  the SDSS database. They use different stellar population synthesis
  models and different methods:
  The table \texttt{stellarMassPCAWiscM11} provides stellar mass
  estimates using the method of \cite{chen:evolution} with the stellar
  population synthesis models of \cite{maraston:stellar}. These are
  the stellar mass estimates used for the plots in
  Fig.\,\ref{fig:sdssmarkcor}.\\
  The table \texttt{stellarMassPCAWiscBC03} provides stellar mass
  estimates using the method of \cite{chen:evolution} with the stellar
  population synthesis models of \cite{bruzual:stellar},
  and the table \texttt{stellarMassStarformingPort} provides stellar
  mass estimates using the method of \cite{maraston:evidence}, see
  also \cite{maraston:stellarmass}.

  Irrespective of the method, $m_\text{st} = \log M_\text{st}/M_\odot$
  is used as a mark in the analysis. $M_\text{st}$ is the stellar mass
  content and $M_\odot$ the solar mass. Both $m_\text{st}$ and the
  magnitude $M_r$ are logarithmic in mass and luminosity respectively.

\item{\textbf{Velocity Dispersion:}} The velocity dispersion
  $\sigma_v$ inside the galaxy is estimated from the spectra as
  described in \cite{bolton:spectral} and is directly read from the
  database view \texttt{SpecObj}.
\end{description}

The volume limited samples comprise galaxies with luminosity distance
$d\le d_\text{lim.}$ and absolute magnitude $M_r\le M_\text{lim.}$.
The limiting absolute magnitude is
$M_\text{lim.}=m_\text{lim.}-D_\text{lim.}$ with the (conservative)
limiting magnitude $m_\text{lim.}=17.7$ and the limiting distance
module $D_\text{lim.}=5\log_{10}(d_\text{lim.}/10\text{pc})$; also
galaxies close by with luminosity distance $d\le50$\,Mpc are
discarded.  Mainly, the volume limited sample with
$d_\text{lim.}=600~\text{Mpc}$ and 201722~galaxies is used, but also
samples with $d_\text{lim.}=300$\,Mpc and $900~\text{Mpc}$ are
considered.

\subsection{Halo samples from the MultiDark simulations}
\label{sec:samples-multidark}

The halo catalogues are constructed from the so called MultiDark
simulations --- dark matter simulations as described in
\cite{prada:haloconcentration} and \cite{klypin:multidark}.  The MDPL2
and BigMDPL simulations have a box size of 1\,Gpc/h and 2.5\,Gpc/h
respectively, with Planck--like cosmology
$\Omega_m=0.307115$, $\Omega_{\Lambda}=0.692885$, $\Omega_\text{rad}=0.0$,
$\omega_0=-1.0$, $h=0.6777$. 
The dark matter halos were identified using the Rockstar halo-finder
\citep{behroozi:rockstar}. These halo samples can be downloaded from
the CosmoSim database \url{https://www.cosmosim.org/} as described in
\cite{riebe:multidark}.
\begin{figure}
\tiny
\begin{verbatim}
select rockstarId, x, y, z, Mvir, spin, axisratio_3_1
from MDPL2.Rockstar 
where snapnum = 125 and Mvir >= 1.0e12 and pId = -1
\end{verbatim}
\normalsize
\caption{The SQL code used on CosmoSim to extract rockstar halos with
  $M_\text{vir}\ge 10^{12}M_\odot/h$ at $z=0$ from the MDPL2
  simulation.   \label{fig:sqlcosmosim} }
\end{figure}
Fig.\,\ref{fig:sqlcosmosim} shows the SQL--code used to extract one of
the desired halo samples from the CosmoSim database.  About
$4\times10^6$ distinct halos
with a virial mass $M_\text{vir}\ge 10^{12}M_\odot/h$ are selected.
With \texttt{snapnum=125} we select the $z=0$ samples and with
\texttt{pId=-1} we ask for distinct halos only.  The virial mass
$M_\text{vir}$, the spin $\lambda$, and the shape $s$ are used as
marks (see below). They can be accessed directly from the database. To
facilitate the calculations of the scale dependent correlation
functions a random subsample comprising 25\% of the halos is used
(about $10^6$ halos). A comparison with the results from 10\% and 50\%
subsampling shows that the results for the scale dependent correlation
coefficients clearly stabilise for 25\% subsampling.
\begin{description}
\item{\textbf{Mass:}} The mass $M_\text{vir}$ within the virial radius
  is calculated from the number of bound particles in the halo. The
  major task of this phase--space halo finder is to reliably assign
  the dark matter particles to a halo, using several steps as detailed
  in \cite{behroozi:rockstar}.
\item{\textbf{Spin:}} The dimensionless spin parameter $\lambda$ is
  used to quantify the rotation of galactic systems (see e.g.\
  \citealt{fall:formation}).
%
%
\item{\textbf{Shape:}} The axial ratios of the mass ellipsoid are
  determined according to the method of \citet{allgood:shape} using
  the eigenvalues of the (reduced) inertia tensor of the halo.  The
  ratio of the smallest ellipsoid axes to the largest ellipsoid axes
  is then used as an overall shape parameter $s$.
\end{description}
To investigate systematic effects also halo samples with mass cuts
$M_\text{vir}\ge 5\times10^{11}M_\odot/h$ and
$M_\text{vir}\ge 10^{13}M_\odot/h$ have been extracted from the MDPL2
and BigMDPL simulations. Also the mass, the spin, and the axes ratios
of the ellipsoidal shape determined from FoF--halos have been used
(see \citealt{riebe:multidark} and \url{https://www.cosmosim.org/} for
details) .

\section{A toy model}
\label{sec:toy}

The following model is a straightforward extension of the marked
Poisson process discussed by \cite{waelder:variograms}.  This model
will serve as an illustration that one is able to unambiguously
extract a scale from a marked point distribution using the scale
dependent correlation coefficient $\textbf{cor}(m_1,m_2\,|\,r)$.  It
is not meant to be a viable model for the galaxy distribution.

One starts with a Poisson process, i.e.\ randomly distributed points
with number density $\varrho$. As suggested by
\cite{waelder:variograms}, one assigns to each point the number of
other points within a radius $R$ as a mark $m_1$. This mark $m_1$ is a
Poisson random variable with the mean mark
$E[m_1]=\overline{m}=\varrho\frac{4\pi}{3}R^3$ and the variance
$\sigma^2_{m_1}=\overline{m}$. Therefore the probability of observing
$n$ points in a sphere with radius $R$ is
$p(n)=\frac{\overline{m}^n}{n!}\mathrm{e}^{-\overline{m}}$.

As an extension of this model, the second mark on a point is slaved to
its first mark by $m_2=m_1$. This construction leads to the covariance
$\text{cov}(m_1,m_2)=\sigma^2_{m_1}$ and perfect overall correlation
$\text{cor}(m_1,m_2)=1$.
For the Poisson point process, one can calculate the desired scale
dependent correlation coefficients easily. The point $\mathbf{x}$ is
marked with $m_2=m_1$ as described above. If a second point at
$\mathbf{y}$ is more distant than $|\mathbf{x}-\mathbf{y}|=r>R$, the
number of points inside the sphere around point $\mathbf{x}$ is
independent from the point at $\mathbf{y}$ and
$\textbf{cov}(m_1,m_2\,|\,r)=\text{cov}(m_1,m_2)$.
Under the condition that the second point at $\mathbf{y}$ is closer
than $R$, at least one point is always in the sphere around the point
at $\mathbf{x}$. Considering a Poisson point process, all the other
points are still independent from this point at $\mathbf{y}$. Now the
probability $q(l)$ of observing $l$ points in the sphere around
$\mathbf{x}$ is $q(0)=0$ and $q(l)=p(l-1)$. This allows us to
calculate
\begin{align*}
\textbf{cov}(m_1,m_2\,|\,r) 
& = E_q[(m_1-\overline{m})(m_2-\overline{m})] \\
&  = E_q\left[(m_1-\overline{m})^2\right] =\overline{m}+1,
\end{align*}
for $r\le R$. $E_q$ is the expectation with respect to the
probabilites $q(l)$.
Joining the results from above one obtains
\begin{equation}
\frac{\textbf{cor}(m_1,m_2\,|\,r)}{\text{cor}(m_1,m_2)} 
= \frac{\textbf{cov}(m_1,m_2\,|\,r)}{\text{cov}(m_1,m_2)} =
\begin{cases}
\frac{\overline{m}+1}{\overline{m}} & \text{for } r\le R,\\
1 & \text{for } r>R.
\end{cases}
\label{eq:markedpoi}
\end{equation}

A similar reasoning allows the calculation of
$\widetilde{\textbf{cor}}(m_1,m_2\,|\,r)$: If the second point at
$\mathbf{y}$ is farther away than $R$, we get
$\overline{m}(r)=\overline{m}$ and $\sigma_m^2(r)=\sigma_m^2$ and
therefore
$\widetilde{\textbf{cov}}(m_1,m_2\,|\,r)=\textbf{cov}(m_1,m_2)$. If
the second point is closer than $R$, one obtains
$\overline{m}(r)=E_q[m_1]$ and
$\sigma_m^2(r)=E_q[(m_1-\overline{m}(r))^2]$, and
\begin{align*}
\widetilde{\textbf{cov}}(m_1,m_2\,|\,r) 
& = E_q[(m_1-\overline{m}(r))(m_2-\overline{m}(r))]\\
& = E_q\left[(m_1-\overline{m}(r))^2\right]  = \sigma_m^2(r).
\end{align*}
Putting everything together
$\widetilde{\textbf{cor}}(m_1,m_2\,|\,r)/\text{cor}(m_1,m_2)\equiv1$
for all radii $r$. The scale $R$ cannot be resolved with
$\widetilde{\textbf{cor}}(m_1,m_2\,|\,r)$.

\begin{figure}
  \centering
  \includegraphics[width=7cm]{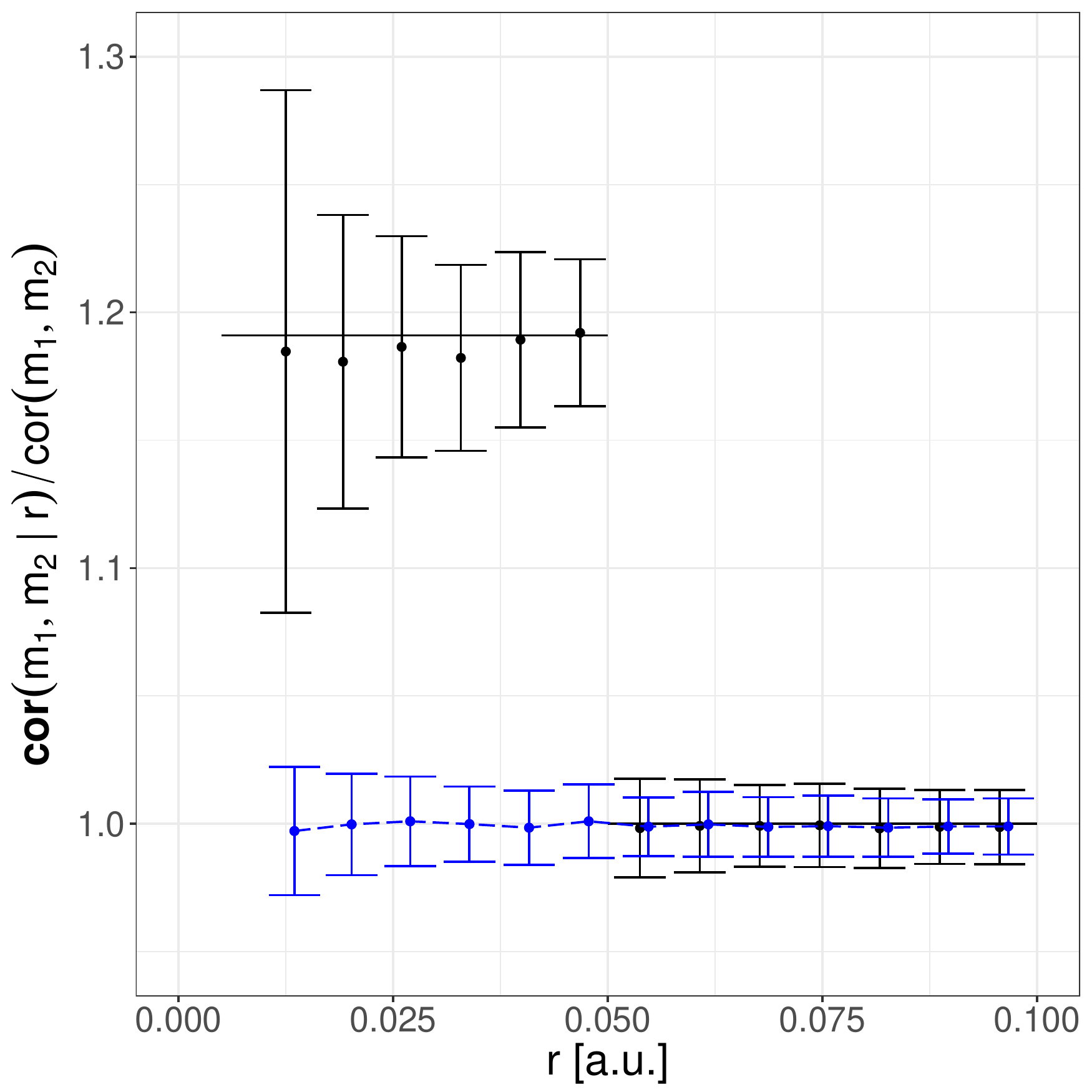}
  \caption{ $\textbf{cor}(m_1,m_2\,|\,r)/\text{cor}(m_1,m_2)$
    estimated from 100~realisations of a marked Poisson process with
    $R=0.05$ and $\varrho=10000$ in the unit box (points with
    one--$\sigma$ error bars). The solid line is the theoretical curve
    according to Eq.\,(\ref{eq:markedpoi}). The blue dashed curve shows
    the results for
    $\widetilde{\textbf{cor}}(m_1,m_2\,|\,r)/\text{cor}(m_1,m_2)$
    (with $x$-coordinates slightly shifted for better visibility).}
  \label{fig:markedpoi}
\end{figure}

In Fig.\,\ref{fig:markedpoi} the estimated
$\textbf{cor}(m_1,m_2\,|\,r)$ for the marked Poisson process is
compared to the theoretical expectation showing perfect agreement. The
jump in $\textbf{cor}(m_1,m_2\,|\,r)$ is resolved, marking the
built-in scale.  As it should be
$\widetilde{\textbf{cor}}(m_1,m_2\,|\,r)/\text{cor}(m_1,m_2)$ is
approximately 1 on all scales.
This simple model illustrates that a built--in scale in the
correlation pattern of the marks can be resolved unambiguously with
$\textbf{cor}(m_1,m_2\,|\,r)$, whereas the alternative definition
$\widetilde{\textbf{cor}}(m_1,m_2\,|\,r)$ does not allow this.

\end{document}